\documentclass[fleqn,12pt]{article}
\usepackage{amsmath}
\usepackage{epsfig}
\usepackage{amsfonts,amssymb,latexsym}

\ifx\href\asklfhas\newcommand{\href}[2]{#2}\fi

\newdimen\squaresize \squaresize=12pt
\newdimen\thickness \thickness=0.7pt

\def\square#1{\hbox{\vrule width \thickness
   \vbox to \squaresize{\hrule height \thickness\vss
      \hbox to \squaresize{\hss#1\hss}
   \vss\hrule height\thickness}
\unskip\vrule width \thickness} \kern-\thickness}

\def\cut#1{\hbox{\vrule width-1 \thickness
   \vbox to \squaresize{\hrule height-1 \thickness\vss
      \hbox to \squaresize{\hss#1\hss}
   \vss\hrule height-1\thickness}
\unskip\vrule width +4 \thickness} \kern-\thickness}

\def\vsquare#1{\vbox{\square{$#1$}}\kern-\thickness}

\makeatletter
\newbox\slashbox \setbox\slashbox=\hbox{$/$}
\def\pFMslash#1{\setbox\@tempboxa=\hbox{$#1$}
  \@tempdima=0.5\wd\slashbox \advance\@tempdima 0.5\wd\@tempboxa
  \copy\slashbox \kern-\@tempdima \box\@tempboxa}

\makeatother

\newcommand{\ft}[2]{{\textstyle {\frac{#1}{#2}} }}

\newcommand{\I}{{\rm i}}

\newcommand{\Beps}{\overline{\Ge}}
\newcommand{\Bchi}{\overline{\chi}}
\newcommand{\Bpsi}{\overline{\psi}}

\newcommand{\be}{\begin{equation}}
\newcommand{\ee}{\end{equation}}
\newcommand{\ben}{\begin{displaymath}}
\newcommand{\een}{\end{displaymath}}
\newcommand{\ba}{\begin{eqnarray}}
\newcommand{\ea}{\end{eqnarray}}

\newcommand{\non}{\nonumber\\}
\newcommand{\bean}{\begin{eqnarray*}}
\newcommand{\eean}{\end{eqnarray*}}

\newcommand{\mathon}{\mathversion{bold}}
\newcommand{\mathoff}{\mathversion{normal}}

\def\moth{\mathsurround=0pt}
\newdimen\zo \zo=0pt

\def\tick{\leaders\hrule height 0.5ex depth 0pt \hskip 0.5pt}
\def\upboxfill{$\moth \setbox\zo\hbox{\tick}%
  \hskip 2pt\hbox to 0pt{$\tick$\hss}\hrulefill \hbox to
6pt{$\tick$\hss}$}

\def\dtick{\leaders\hrule height .34pt depth .5ex \hskip 0.5pt}
\def\downboxfill{$\moth \setbox\zo\hbox{\dtick}%
  \hskip 2pt\hbox to 0pt{$\dtick$\hss}\hrulefill \hbox to
6pt{$\dtick$\hss}$}

\newcommand{\la}{\label}

\newcommand{\Ref}[1]{(\ref{#1})}

\makeatletter \@addtoreset{equation}{section} \makeatother

\def\be{\begin{equation}}
\def\ee{\end{equation}}
\def\bea{\begin{eqnarray}}
\def\eea{\end{eqnarray}}

\newcommand{\vl}{{\vphantom{[}}}

\newcommand{\equ}{\!=\!}

\newcommand{\Ga}{\alpha}

\newcommand{\Ge}{\epsilon}

\newcommand{\Gg}{\gamma}
\newcommand{\GG}{\Gamma}

\newcommand{\cV}{{\cal V}}

\newcommand{\dA}{{\dot{A}}}

\newcommand{\VV}[2]{{\cV^{\,#1}{}\!^\vl_{#2}}}

\begin{document}

\begin{titlepage}

\begin{flushright}\small 
\end{flushright}
%
\vskip 8mm
\mathon
{\LARGE {\bf On The Supersymmetric Solutions of 
  $D=3$\\[1ex] 
  Half-maximal Supergravities}}
  \mathoff
\vskip 14mm

\noindent
{{\bf Nihat Sadik Deger}\\ 
{\small\em 
\phantom{xxx}Department of Mathematics,
Bogazici University, \\[-.3ex]
\phantom{xxx}34342, Bebek, Istanbul, Turkey\\[-.3ex]
\phantom{xxx}{\tt sadik.deger@boun.edu.tr}}}
\vskip 3mm

\noindent
{{\bf Henning Samtleben}\\
{\small\em
\phantom{xxx}Universit\'e de Lyon, Laboratoire de Physique, UMR 5672, CNRS,\\[-.3ex]
\phantom{xxx}Ecole Normale Sup\'erieure de Lyon,
46 all\'ee d'Italie, F-69364 Lyon cedex 07, France \\[-.3ex]
\phantom{xxx}{\tt henning.samtleben@ens-lyon.fr}}}
\vskip 3mm

\noindent
{{\bf \"{O}zg\"{u}r Sar{\i}o\u{g}lu}\\
{\small\em
\phantom{xxx}Department of Physics,
Middle East Technical University, \\[-.3ex] 
\phantom{xxx}06531, Ankara, Turkey  \\[-.3ex]
\phantom{xxx}{\tt sarioglu@metu.edu.tr}}}
\vskip 4mm

\vskip15mm

\begin{center} {\bf Abstract } 
\end{center}
\begin{quotation}\noindent
We initiate a systematic study of the solutions of three-dimensional matter-coupled
half-maximal ($N=8$) supergravities which admit a Killing spinor.
To this end we analyze in detail the invariant tensors built from spinor bilinears,
a technique originally developed and applied in higher dimensions.
This reveals an intriguing interplay with the scalar target space geometry
$SO(8,n)/(SO(8)\times SO(n))$.
Another interesting feature of the three-dimensional case is the 
implementation of the duality between vector and scalar fields in this framework.
For the ungauged theory with timelike Killing vector, 
we explicitly determine the scalar current and show that
its integrability relation reduces to a covariant holomorphicity equation,
for which we present a number of explicit solutions.
For the case of a null Killing vector, we give the most general solution
which is of pp-wave type.

\end{quotation}
\end{titlepage}
\eject  

\tableofcontents

\bigskip
\bigskip

\section{Introduction}
\setcounter{equation}{0}
\label{sec:introduction}

Classical solutions of supersymmetric gauge theories and supergravities that preserve some fraction
of supersymmetry play a distinguished role in these theories as they typically exhibit particular
stability and non-renormalization properties, due to the rigidness of the underlying supersymmetry
algebra and its representations.
Moreover, in the search of new solutions to supersymmetric theories it is often technically simpler to solve
the first order Killing spinor equations rather than to address the full second order field equations.
The systematic study of these questions in supergravity theories
goes back to the seminal work of Tod~\cite{Tod:1983pm,Tod:1995jf}
(see~\cite{Gibbons:1982fy} for earlier work),
and has further and systematically 
been developed in~\cite{Gauntlett:2002sc,Gauntlett:2002nw} and a large body of follow-up work
for higher-dimensional supergravity theories.

In this approach, one assumes the existence of a Killing spinor and considers all bilinear tensors 
that can be constructed from it. The Killing spinor equations together with standard Fierz identities translate into 
a set of algebraic and differential identities among these bilinear tensors which can be employed to constrain
the structure of the geometry and the matter dynamics. In many cases this allows for a complete classification
of the supersymmetric solutions of the theory.
So far, this bilinear tensor analysis has mainly been employed and proven very useful 
to construct and classify supersymmetric solutions in theories
with eight real supercharges, such as $N=2$ supergravity in four dimensions, see e.g.~\cite{Caldarelli:2003pb,Meessen:2006tu,Huebscher:2006mr,Cacciatori:2008ek,Klemm:2009uw,Hristov:2009uj}. 
In the context of half-maximal supergravities (i.e.\ sixteen real supercharges), 
the analysis becomes more involved due to the richer structure of the extended $R$-symmetry groups.
Previous work on such theories includes

\begin{itemize}

\item[(i)]
the pure ungauged $D=4, N=4$ supergravity~\cite{Bellorin:2005zc},

\item[(ii)]
the pure ungauged $D=5, N=4$ supergravity~\cite{Liu:2006is},

\item[(iii)]
the pure ungauged and the $SU(2)$ gauged $D=7, N=1$ supergravity~\cite{Cariglia:2004qi,MacConamhna:2004fb},

\end{itemize}

\noindent
but has throughout been restricted to theories without additional vector multiplets, 
leading in particular to scalar target spaces of very small dimensions ($\le 2$).

In this paper, we apply the bilinear tensor techniques to the matter-coupled half-maximal $N=8$ supergravity 
in three dimensions~\cite{Marcus:1983hb,Nicolai:2001ac}.
The three-dimensional case exhibits a number of interesting properties: 
in particular, the metric in three dimensions does not carry propagating degrees of freedom,
and vector gauge fields can be dualized into scalar fields.
As a result, the entire dynamics of these theories takes place in 
the scalar sector, whose target space is a coset space manifold
$SO(8,n)/(SO(8)\times SO(n))$ of dimension~$8n$.
The bilinear tensor analysis which we present in this paper leads to
an interesting interplay of the Killing spinor equations and this target space geometry.
The structures we exhibit in this paper represent a first example of the structures
that will also appear in higher dimensions upon inclusion of vector multiplets with 
coset space geometry. The advantage of the three-dimensional setting is the fact that
all matter dynamics is uniformly described in the scalar sector.

Upon passing to the gauged $N=8$ supergravity, vector fields have to be introduced in
three dimensions, however these do not represent additional propagating degrees of freedom,
but enter with a Chern-Simons coupling and are related to the scalar fields by
(a non-abelian version of) their standard first order duality equations.
As such, the structure of the theory (and its Killing spinor equations) is still
organized by the full group $SO(8,n)$ even though the explicit choice of a gauge group
breaks this global symmetry of the theory.
We will analyze the structure of the Killing spinor equations and their consequences for the
matter fields in a fully $SO(8,n)$ covariant manner without specifying the three-dimensional gauge group.
Of course, when further exploiting these structures to construct explicit solutions, this choice 
will have to be made.

Some three-dimensional supersymmetric solutions of the half-maximal ungauged theory 
with higher-dimensional origin in heterotic string theory
(compactified on a seven torus)
have been constructed in \cite{Forste:1997ac,Bakas:1997jy,Bourdeau:1997gm}, see also
\cite{Greene:1989ya,Sen:1994wr} for earlier work. 
Some supersymmetric solutions of the gauged $N=8$
theory have been identified in~\cite{Berg:2001ty,Berg:2006ng,Gava:2010rx},
among them are analytic domain wall solutions which can be interpreted as holographic RG flows
in particular scenarios of AdS$_3$/CFT$_2$ dualities.
In all these different contexts it would clearly be interesting and highly desirable to dispose of a more systematic
approach to construct and classify the possible supersymmetric solutions.

In this paper, we take a first step in this direction and apply the bilinear tensor techniques to the 
matter-coupled $D=3$, $N=8$ supergravity, which is reviewed in section~2.
We construct in section 3 the bilinear tensors from the Killing spinor and translate the full 
content of the Killing spinor equations into a set of algebraic and differential identities among these tensors. 
As usual, these tensors provide a timelike or null Killing vector that
generates a symmetry of the full solution (i.e.\ including the matter sector).
In the timelike case, the $SO(8)$ $R$-symmetry breaks down to $SO(2)\times SO(6)$ and 
we show in section~4 that the Killing vector can be extended to a canonical tetrad in which we expand 
the fields and obtain the
general solution for the scalar current. The $8n$ scalar fields naturally split into the $4n+4n$
eigenvectors of an antisymmetric matrix $\Omega$ related to the $SO(2)$ part of the 
composite connection on the coset space manifold.
With the general solution for the scalar current we analyze the first order duality equation between vector
and scalar fields and show that it reduces to a two-dimensional equation for the field-strength of
the undetermined part of the vector fields. In particular, for the ungauged theory we obtain an explicit
expression of the vector fields in terms of the scalar fields of the theory.

In section 4.4, we give a general analysis to determine to which extent the equations of
motion are satisfied as a consequence of the existence of a Killing spinor.
In order to construct explicit solutions, the Killing spinor equations have to be amended by the
integrability relations for the scalar current and we do this in section 5 for the ungauged theory,
leaving the full analysis of the gauged theory for future work.
In this case, the three-dimensional metric takes diagonal form with the conformal factor
satisfying a Liouville-type equation. 
The integrability relations for the scalar current reduce to a covariant holomorphicity equation 
(with the composite connection of the scalar target space) 
for its unknown component. As an illustration, we employ a simple 
ansatz to derive in section 5 explicit solutions with $1\le n\le4$ (active) matter multiplets.
Finally, in section~6 we analyze the case of a null Killing vector in the ungauged theory
for which we derive the most general solution for metric and scalar fields which is a pp-wave.
Appendix~A collects our $SO(8,n)$ conventions and some useful formulae.

\section{Half-maximal supergravities
in three dimensions}

In this section we review the structure of half-maximal (i.e.\ $N=8$)
supergravity theories in three dimensions. In three dimensions, pure (super-)gravity
does not possess propagating degrees of freedom. Moreover, vector fields
can be dualized into scalar fields, such that all propagating degrees of freedom 
may be accommodated in the scalar sector of the theory.
The ungauged theory is described by an $SO(8,n)/(SO(8)\times SO(n))$ 
coset space sigma-model coupled to gravity~\cite{Marcus:1983hb}.
The general gauged theory in which gauge fields are added with a Chern-Simons coupling
has been constructed in~\cite{Nicolai:2001ac}, to which we refer for details and conventions.

\subsection{Bosonic field content and Lagrangian}

The Lagrangian of the general gauged $N=8$ supergravity in three dimensions
is given by \cite{Nicolai:2001ac}
\bea
{\cal L}&=& -\ft14 \,eR
+ {\cal L}_{\rm CS} + 
\ft14 e \,{\cal P}^{\mu\,Ir} {\cal P}_\mu^{\,Ir} 
+{\cal L}_{\rm pot} + {\cal L}_{\rm F}\;.
\label{L}
\eea
Let us explain each term separately.
We use signature $(+--)$, and denote the dreibein determinant by $e=\sqrt{\det\,g_{\mu\nu}}$,
such that the first term is the standard Einstein-Hilbert term of three-dimensional gravity.

The vector fields in three-dimensional supergravity do not represent propagating degrees
of freedom, but couple in the gauged theory 
with a Chern-Simons kinetic term ${\cal L}_{\rm CS}$ given in (\ref{LCS}) below,
such that they are related by
their first order equations of motion to the scalar fields. 
The gauge group $G_0$ is a subgroup of $SO(8,n)$, the isometry group of the scalar sector.
Its generators $\Xi_{\cal MN}$ can be represented as linear combinations of the $SO(8,n)$ generators
$X^{\cal MN}=-X^{\cal NM}$ by means of the {\em embedding tensor} $\theta_{\cal MNKL}$ as
\bea
\Xi_{\cal MN} &\equiv&  \theta_{\cal MNKL}\, X^{\cal KL}
\;,
\label{generators}
\eea
where indices {\small ${\cal M}, {\cal N}, \dots$} label the vector representation of $SO(8,n)$.
They can be raised and lowered with the $SO(8,n)$ invariant indefinite metric $\eta_{\cal MN}$.
Moreover, we split these indices into $\{{\small{\cal M}}\} \rightarrow \{ I,r \}$ according to the signature of $\eta_{\cal MN}$,
i.e.\ $I = 1, \dots, 8$, $r=1, \dots, n$,
with $I$ labeling the vector representation ${\bf 8}_v$ of $SO(8)$.
The specific form of the gauge group $G_0$, its dimension and its embedding into $SO(8,n)$ are entirely encoded
in the constant tensor $\theta_{\cal MNKL}$. Supersymmetry requires $\theta_{\cal MNKL}$ to be completely
antisymmetric in its four indices.\footnote{In fact, the condition which supersymmetry imposes is slightly weaker,
it also allows for contributions of the type 
$\theta_{\cal MNKL}=\eta_{{\cal M}[{\cal K}}\xi_{{\cal L}]{\cal N}}-\eta_{{\cal N}[{\cal K}}\xi_{{\cal L}]{\cal M}}$
with a symmetric $\xi_{\cal MN}=\xi_{\cal NM}$, cf.~\cite{deWit:2003ja}, but we shall not consider these cases here.}
In addition, gauge covariance requires $\theta_{\cal MNKL}$ to satisfy the bilinear relation
\bea
2\theta^{\cal Q}{}_{\cal KL[M}\theta_{\cal N]RSQ}&=&
\theta^{\cal Q}{}_{\cal RMN}\theta_{\cal KLSQ} - \theta^{\cal Q}{}_{\cal KLR}\theta_{\cal MNSQ}
\;.
\label{quadcon}
\eea

A standard example of gauged ${\cal N}=8$ supergravity corresponds to the choice
\be
\theta_{IJKL} ~=~ \left\{
\begin{array}{rl}
\epsilon_{IJKL} & \mbox{for}\quad  I,J,K,L \in \{1,2,3,4\} \\
\alpha \, \epsilon_{IJKL} & \mbox{for} \quad I,J,K,L \in \{5,6,7,8\} \\[.5ex]
0 & \mbox{otherwise}
\end{array}
\right. \;,
\label{theta}
\ee
which solves (\ref{quadcon})
and describes the gauging of an $SO(4)\times SO(4) \subset SO(8)$ subgroup of $SO(8,n)$,
the parameter $\alpha$ corresponding to  the ratio of coupling constants of
the two $SO(4)$ factors. This theory describes the coupling of the spin $1/2$ multiplet in the reduction of
six-dimensional supergravity on $AdS_3\times S^3$ (for $\alpha=0$), or $AdS_3\times S^3\times S^3$.
More complicated gaugings have been given in \cite{Nicolai:2003ux,Hohm:2005ui}
which describe the couplings of higher massive multiplets in this compactification.
In the following we shall not specify the explicit gauge group, 
but derive the constraints on the structure of supersymmetric solutions 
for general choice of $\theta_{\cal MNKL}$.

Explicitly, the Chern-Simons term ${\cal L}_{\rm CS}$ is given by\footnote{
Here and in the following, we denote by $\varepsilon_{\mu\nu\rho}$ the
totally antisymmetric spacetime tensor, i.e.\ in particular it carries
the determinant $e$ of the vielbein: $\varepsilon_{123}=e$, etc.}
\be
{\cal L}_{\rm CS} ~=~
-\ft14\,e\,\varepsilon^{\mu\nu\rho}\,g\theta_{\cal KLMN}\,A_\mu{}^{\cal KL}
\Big(\partial_\nu A_\rho\,{}^{\cal MN}
+\ft83\,g\, \eta^{\cal MR}\,\theta_{\cal RSPQ}\, 
A_\nu{}^{\cal PQ} A_\rho{}^{\cal SN} \Big) \;,
\label{LCS}
\ee
where the parameter $g$ denotes the gauge
coupling constant.

The scalar sector in the Lagrangian \Ref{L} is described by a gauged coset space sigma-model
$SO(8,n)/(SO(8)\times SO(n))$.
It can be parametrized by a group-valued $SO(8,n)$ matrix ${\cal S}$
(i.e.\ a matrix satisfying ${\cal S}\eta {\cal S}^T=\eta$)
which defines the 
left-invariant scalar currents as
\bea
{\cal J}_\mu~\equiv~
{\cal S}^{-1}{\cal D}_\mu {\cal S} &\equiv&  
{\cal S}^{-1} \,\Big(\partial_\mu +   
g \theta_{\cal KLMN} \,A_\mu{}^{\cal KL} X^{\cal MN}\Big)\, {\cal S}
~\in~{\rm Lie}\,SO(8,n)
\;,
\label{current}
\eea
with the covariant derivatives carrying the gauge group generators $\Xi_{\cal KL}$ of (\ref{generators}).
We split these currents according to
\bea
{\cal Q}_\mu^{IJ} ~\equiv ~ ({\cal J}_\mu)^{IJ} 
\;,
\quad
{\cal Q}_\mu^{rs} ~\equiv ~ ({\cal J}_\mu)^{rs} 
\;,
\quad
{\cal P}_\mu^{Ir} ~\equiv ~ ({\cal J}_\mu)^{Ir} 
\;,
\label{defPQ}
\eea
corresponding to the compact generators  $X^{IJ}$, $X^{rs}$ and noncompact generators $Y^{Ir}$
of the algebra $\mathfrak{so}(8,n)\equiv{\rm Lie}\, SO(8,n)$, see appendix~\ref{app:so} for our algebra conventions. 
We may explicitly separate the gauge field contributions to the currents as
\bea
{\cal Q}_\mu^{IJ}&=& 
Q_\mu^{IJ}+2gA_\mu{}^{\cal KL}{\cal V}^{\cal MN}{}_{IJ}\,\theta_{\cal KLMN}
\;,
\nonumber\\
{\cal P}_\mu^{Jr}
  &=&
 {P}_\mu^{Jr}
+g A_\mu^{\cal KL} {\cal V}^{\cal MN}{}_{Jr}\,\theta_{\cal KLMN}
\;,
\label{PgP}
\eea
where we use the short-hand notation 
${\cal V}^{\cal MN}{}_{{\cal PQ}}\equiv 2{\cal S}^{[{\cal M}}{}_{\cal P}{\cal S}^{{\cal N}]}{}_{\cal Q}$,
and $Q^{IJ}_\mu \equiv {\cal Q}^{IJ}_\mu |_{g=0}$, $P^{Ir}_\mu \equiv {\cal P}^{Ir}_\mu |_{g=0}$ are
defined by (\ref{current}) at $g=0$,
i.e.\ represent the currents of the ungauged sigma-model, cf.~appendix~A.
The scalar kinetic term in (\ref{L}) is defined in terms of the noncompact components ${\cal P}_\mu^{Jr}$
of the scalar current and thus invariant under local transformations
\bea
{\cal S} ~\rightarrow~ {\cal S}\,H(x)\;,
\label{coset}
\eea
with $H(x)\in SO(8)\times SO(n)$,
taking care of the coset redundancy.
The integrability equations
$2\partial_{[\mu}{\cal J}_{\nu]}+[{\cal J}_\mu,{\cal J}_\nu]=
{\cal S}^{-1} [{\cal D}_{\mu},{\cal D}_{\nu}] {\cal S}$
induced by the definition of (\ref{current}), translate into
\bea
{\cal D}^{\vphantom{I}}_{[\mu} {\cal P}^{Ir}_{\nu]} &\equiv&
\partial^{\vphantom{I}}_{[\mu} {\cal P}^{Ir}_{\nu]}+{\cal Q}_{[\mu}^{IJ} {\cal P}^{Jr}_{\nu]}
+{\cal Q}_{[\mu}^{rs} {\cal P}^{Is}_{\nu]}
~=~
\ft12\,
g\theta_{\cal KLMN}\,{\cal F}_{\mu\nu}{}^{\cal KL} {\cal V}^{\cal MN}{}_{Ir}
\;,
\label{intP}\\[.5ex]
{\cal Q}^{IJ}_{\mu\nu} &\equiv& 2\partial_{[\mu}{\cal Q}^{IJ}_{\nu]} + 2 {\cal Q}^{IK}_{[\mu}{\cal Q}^{KJ}_{\nu]}
~=~ -2{\cal P}_{[\mu}^{Ir}{\cal P}_{\nu]}^{Jr}
+g\theta_{\cal KLMN}\,{\cal F}_{\mu\nu}{}^{\cal KL} {\cal V}^{\cal MN}{}_{IJ}
\;,\qquad\quad
\label{intQ1}\\[.5ex]
{\cal Q}^{rs}_{\mu\nu} &\equiv& 2\partial_{[\mu}{\cal Q}^{rs}_{\nu]} + 2 {\cal Q}^{rt}_{[\mu}{\cal Q}^{ts}_{\nu]}
~=~ -2{\cal P}_{[\mu}^{Ir}{\cal P}_{\nu]}^{Is}
+g\theta_{\cal KLMN}\,{\cal F}_{\mu\nu}{}^{\cal KL} {\cal V}^{\cal MN}{}_{rs}
\;,
\label{intQ2}
\eea
with the nonabelian field strength ${\cal F}_{\mu\nu}{}^{\cal KL}$\,.

Finally, the gauged theory carries a scalar potential whose form is fully determined by
supersymmetry as
\be
{\cal L}_{\rm pot} ~=~
-eg^2\,{V}\;,\qquad
\mbox{with}\quad
V~\equiv~ 
\ft18\,A_2^{A\dot{A} r}A_2^{A\dot{A} r}-\ft14 A_1^{AB}A_1^{AB}  \;,
\label{potential}
\ee
in terms of the scalar $SO(8)$ tensors $A_1$, $A_2$, defined by
contracting the embedding tensor of (\ref{generators}) with the scalar fields
as
\ba
A_{1\, AB}&=& 
-\ft1{48}\,\Gamma^{IJKL}_{AB}\,
{\cal V}^{\cal KL}{}_{IJ}\,{\cal V}^{\cal MN}{}_{KL}\,\theta_{\cal KLMN}
\;,
\non
A_{2\, A\dot{A} r}&=&
-\ft1{12}\,\Gamma^{IJK}_{A\dot{A}}\,
{\cal V}^{\cal KL}{}_{IJ}\,{\cal V}^{\cal MN}{}_{Kr}\,\theta_{\cal KLMN} \;.
\label{A1A2}
\ea
Here $A=1, \dots, 8$, and $\dot{A}=1, \dots, 8$, label the two $SO(8)$
spinor representations ${\bf 8}_s$, ${\bf 8}_c$, and $\Gamma^I_{A\dot{A}}, \Gamma^{IJ}_{AB}$, etc.\ denote
the corresponding $\Gamma$-matrices.
The tensors (\ref{A1A2}) describe the  Yukawa couplings in the fermionic sector
that we describe explicitly in the next subsection.

Up to fermionic contributions, the bosonic equations of motion derived from the Lagrangian (\ref{L}) 
are given by
\bea
R_{\mu\nu}-\ft12 R g_{\mu\nu} &=&
{\cal P}^{Ir}_\mu   {\cal P}^{Ir}_\nu - \ft12 g_{\mu\nu}\,  {\cal P}^{Ir}_\rho   {\cal P}^{\rho\,Ir}
+ 2 g^2 V g_{\mu\nu}
\;,
\label{eqm:Einstein}
\\[1ex]
{\cal D}^\mu\,{\cal P}^{Ir}_\mu &=& 
\ft14\,g^2 \Gamma^{I}_{A\dot{A}} \left( 
A^{\dot{A}r\,\dot{B}s}_3A_2^{A\dot{B}s}
-3A_1^{AB} A_2^{B\dot{A}r}  \right)
\;,
\label{eqm:scalars}
\eea
for the metric and the scalar fields, with the scalar tensor $A^{\dot{A}r\,\dot{B}s}_3$ defined in
(\ref{A3}) below.
Varying the vector fields in the Lagrangian (\ref{L}) gives rise to
the first order duality equation
\bea
  \theta_{\cal KLMN}\, {\cal F}_{\mu\nu}{}^{\cal MN} &=&
\varepsilon_{\mu\nu\rho} \theta_{\cal KLMN} {\cal V}^{\cal MN}{}_{Ir}\,{\cal P}^{\rho\,Ir}\;,
\label{duality}
\eea
which manifests the fact that the vector fields
do not carry propagating degrees of freedom. Note further, that even though 
formally we have introduced ${\rm dim}\,SO(8,n)$ vector fields $A_\mu{}^{\cal MN}$,
only their projections $\theta_{\cal KLMN}\, {A}_{\mu}{}^{\cal MN}$  appear in the equations.
For example, with (\ref{theta}) only 12 vector fields appear in the action (or 6 if $\alpha=0$).

\subsection{Fermions and Killing spinor equations}

The fermionic couplings of the Lagrangian (\ref{L}) are given by
\bea
e^{-1}{\cal L}_{\rm F}&=& 
\ft12 \varepsilon^{\mu\nu\rho} \Bpsi{}^A_\mu {\cal D}_\nu\psi^A_\rho 
-\ft12\,\mathrm{i} \Bchi^{\dot{A} r}\gamma^\mu {\cal D}_\mu\chi^{\dot{A} r}
- \ft12 {\cal P}_\mu^{Ir}\Bchi^{\dot{A} r} 
\Gamma^I_{A\dot{A}}\gamma^\nu\gamma^\mu\psi^A_\nu
\nonumber\\[1ex]
&&{}
+\ft12 g A_1^{AB} \,\Bpsi{}^A_\mu \gamma^{\mu\nu} \psi^B_\nu
+\I g A_2^{A\dot{A} r} \,\Bchi^{\dot{A} r} \gamma^{\mu} \psi^A_\mu
+\ft12 gA_3^{\dot{A} r\, \dot{B} s}\, \Bchi^{\dot{A} r}\chi^{\dot{B} s} 
\;.
\label{LF}
\eea
Here $\chi^{\dot{A} r}$ and $\psi^A_\mu$ are two-component Majorana spinors,
transforming in the ${\bf 8}_c$ and ${\bf 8}_s$ of $SO(8)$, respectively.
Covariant derivatives ${\cal D}_\mu$ on these spinors include the spin connection, the
$SO(8)$ connection
${\cal Q}_{\mu\,\dot{A}\dot{B}}\equiv\ft14\,{\cal Q}_{\mu\,IJ}\,\Gamma^{IJ}_{\dot{A}\dot{B}}$
and
${\cal Q}_{\mu\,AB}\equiv\ft14\,{\cal Q}_{\mu\,IJ}\,\Gamma^{IJ}_{AB}$,
respectively, and the $SO(n)$ connection ${\cal Q}_{\mu}^{rs}$ obtained from (\ref{defPQ}).
We use $\gamma^{\mu\nu\rho}=-\mathrm{i}\varepsilon^{\mu\nu\rho}$
for the three-dimensional $\gamma$-matrices.
The Yukawa-type couplings are given by the scalars tensors $A_1$, $A_2$ from (\ref{A1A2})
and
\bea
A_3^{\dot{A} r\, \dot{B} s} =
\ft1{48}\,\delta^{rs}\,\Gamma^{IJKL}_{\dot{A}\dot{B}}\,
{\cal V}^{\cal KL}{}_{IJ}\,{\cal V}^{\cal MN}{}_{KL}\,\theta_{\cal KLMN}
+\ft1{2}\,\Gamma^{IJ}_{\dot{A}\dot{B}}\,
{\cal V}^{\cal KL}{}_{IJ}\,{\cal V}^{\cal MN}{}_{rs}\,\theta_{\cal KLMN} \;,
\label{A3}
\eea
which also appears in the bosonic equations of motion (\ref{eqm:scalars}).
Combining their definition with (\ref{dV}) shows that
these tensors are related by the differential relations
\bea
{\cal D}_\mu A_1^{AB}&=&
\ft12\left(\Gamma^I_{A\dot{A}}A_{2}^{B\dot{A}r}+
\Gamma^I_{B\dot{A}}A_{2}^{A\dot{A}r}\right)\,{\cal P}^{Ir}_\mu
\;,
\nonumber\\
{\cal D}_\mu A_{2}^{A\dot{A} r} &=&
\ft12\Gamma^I_{B\dot{A}} A_1^{AB}\,{\cal P}_\mu^{Ir}+
\ft12\Gamma^I_{A\dot{B}} A_3^{\dot{A}r\,\dot{B}s}\,{\cal P}_\mu^{Is}+
\ft1{16} \Gamma^J_{A\dot{A}}\Gamma^{IJ}_{\dot{B}\dot{C}}A_3^{\dot{B}r\,\dot{C}s}\,{\cal P}_\mu^{Is}
\;,
\eea
which play an important role in proving supersymmetry of the action (\ref{L}).
The quadratic constraints (\ref{quadcon}) translate into 
various bilinear identities among the scalar tensors, such
as the {supersymmetric Ward identity}
\bea
A_1^{AC}A_1^{BC}-\ft12\,A_2^{A\dot{A} r}A_2^{B\dot{A} r}
&=& \ft18\delta^{AB}\,
\left(
A_1^{CD}A_1^{CD}-\ft12\,A_2^{C\dot{A} r}A_2^{C\dot{A} r}
\right)
\;.
\label{qc1}
\eea

The full action (\ref{L}) is $N=8$ supersymmetric. The Killing spinor equations of the theory follow as usual
by imposing the vanishing of the fermionic supersymmetry
variations on a given bosonic  background
\begin{eqnarray}
  \label{KS1}
 0 &\equiv& \delta_\epsilon \psi_\mu^A ~=~
     (\partial_\mu + \ft14 \omega_\mu{}^{ab}\gamma_{ab})\,\epsilon^A
     +{\cal Q}_{\mu\,AB} \,\epsilon^B 
+ \mathrm{i} g A_1^{AB}\gamma_\mu\,\epsilon^B
\;,
\\[1ex]
  0  &\equiv&{} \delta_\epsilon \chi^{\dot{A} r} ~=~
  \ft{\mathrm{i}}{2}\Gamma^I_{A\dot{A}}\gamma^\mu {\cal P}_\mu^{Ir}\epsilon^A 
  +g A_2^{A\dot{A} r}\epsilon^A
  \;,
  \label{KS2}
\end{eqnarray}
with the scalar tensors $A_{1}$, $A_{2}$ from (\ref{A1A2}) above.
In the rest of the paper we will analyze the consequences of these
equations for supersymmetric solutions of the theory.

\section{Killing spinor bilinears}

We will study in this paper the structure of supersymmetric solutions of the $N=8$ 
theory defined by (\ref{L}).
Let us assume the existence of $p$ (commuting) Killing spinors 
$\epsilon^{A}_{(\alpha)}$, $\alpha=1, \dots, p$, satisfying equations (\ref{KS1}) and (\ref{KS2}). 
From these we can define real scalar and 
vector functions bilinear in the Killing spinors, as follows
\bea
F^{\,AB}_{\alpha\beta} ~\equiv~
\bar\epsilon^{A}_{(\alpha)}\,\epsilon^{B}_{(\beta)}
\;,\qquad
V_{\mu}{}^{\,AB}_{\,\,\alpha\beta} ~\equiv~
\mathrm{i}\,\bar\epsilon^{A}_{(\alpha)}\,\gamma_{\mu}\,\epsilon^{B}_{(\beta)}
\;.
\eea
By construction they satisfy
\bea
F^{AB}_{\alpha\beta} ~=~- F^{BA}_{\beta\alpha}
\;,\qquad 
V_{\mu}{}^{AB}_{\,\,\alpha\beta}~=~  V_{\mu}{}^{BA}_{\,\,\beta\alpha} 
\;.
\eea
In the following, we focus on the case $p=1$ and consider 
the real tensors built from a single commuting Killing spinor $\epsilon^{A}$
\bea
F^{[AB]}~=~\bar\epsilon^{A}\epsilon^{B}\;,\qquad
V_{\mu}{}^{(AB)}~=~\mathrm{i}\,\bar\epsilon^{A}\gamma_{\mu}\,\epsilon^{B}\;.
\label{FV}
\eea
Furthermore, we define the ${\rm SO}(8)$ invariant real combination 
$V_{\mu}\equiv V_{\mu}{}^{AA}$\,.
In the rest of this section we will translate the full content of
the Killing spinor equations (\ref{KS1}), (\ref{KS2}) into a set of algebraic and differential 
relations for the invariant tensors $F^{AB}$ and $V_\mu{}^{AB}$\,.

\subsection{Algebraic relations}
\label{subsec:algeq}

{}The cubic Fierz identities for the Killing spinor $\epsilon^{A}$, together with 
the definitions (\ref{FV}) induce the relations
\bea
V_{\mu}{}^{AB}\gamma^{\mu}\epsilon^{C} &=& -2\mathrm{i} F^{C(A}\epsilon^{B)}
\;,
\nonumber\\
V_\mu{}^{AB}\epsilon^C
-\mathrm{i} \varepsilon_{\mu\nu\rho} V^{\nu\,AB}\gamma^\rho\epsilon^{B}
&=&
-2\mathrm{i} F^{C(A}\gamma_\mu\epsilon^{B)}
\;.
\eea
Similarly, evaluating quartic Fierz identities, one obtains the identities
\bea
V^{\mu}V_{\mu} &=& -2 V^{\mu\,AB}V_\mu{}^{AB} ~=~
2F^{AB}F^{AB} 
\;,
\nonumber\\
V^{\mu\,AB}V_\mu{}^{CD} &=& -2 F^{A(C}F^{D)B}
\;,
\nonumber\\
V_{[\mu}{}^{AB}V_{\nu]}{}^{CD} &=&
\ft12\,\varepsilon_{\mu\nu\rho}\,
\Big(V^{\rho\,A(C}\,F^{D)B}+V^{\rho\,B(C}\,F^{D)A}\Big)
\;,
\nonumber\\
V_{(\mu}{}^{AC}V_{\nu)}{}^{BC}
&=&
V_{(\mu}{}V_{\nu)}{}^{AB} -\ft12 g_{\mu\nu}\,V^{\rho}{}V_{\rho}{}^{AB} 
\;,
\nonumber\\
F^{C[A}\,V_\mu{}^{B]C}&=& -\ft12\,F^{AB}\,V_\mu
\;,
\label{quartic1}
\eea
bilinear in the tensors $F^{AB}$, $V_\mu{}^{AB}$.
From a sextic Fierz identity, one finally finds
\bea
F^{AB}F^{BC}F^{CD} + f^2 \,F^{AD} &=& 0
\;,
\label{sextic}
\eea
with $f^2=\frac12 F^{AB}F^{AB}$\,.
This implies that the antisymmetric matrix $F^{AB}$ has only two nonvanishing eigenvalues $\pm \mathrm{i}f$\,.
Accordingly, it will often be convenient to
change to an explicit basis for the $SO(8)$ spinor indices 
\bea
A=(a,\tilde{a})\;,\qquad a=1,2\;,\quad \tilde{a}=3,\dots,8\;,
\label{basis}
\eea
in which ${F}^{ab}=f\epsilon^{ab}$\,, $F^{a\tilde{a}}=0=F^{\tilde{a}\tilde{b}}$.\footnote{We use conventions $\epsilon^{12}=\epsilon_{12}=1$.}
For non-vanishing $f$, this corresponds to a breaking of the $R$-symmetry
according to $SO(8)\rightarrow SO(2)\times SO(6)$, under which the fundamental representations
branch as
\bea
{\bf 8}_s \rightarrow 1_{-1}+6_0+1_{+1}\;,\quad
{\bf 8}_v \rightarrow  4_{-1/2}+\overline{4}_{+1/2}\;,\quad
{\bf 8}_c \rightarrow  \overline{4}_{-1/2}+4_{+1/2}
\;,
\label{branching}
\eea
with subscripts denoting $SO(2)$ charges.
The second and the last equation of~\Ref{quartic1} then imply 
that the only nonvanishing components of $V_{\mu}{}^{AB}$ are the~$V_{\mu}{}^{ab}$.
The third equation of~\Ref{quartic1} shows that the
three vectors $V_\mu{}^{11}$, $V_\mu{}^{12}= V_{\mu}^{21}$, and $V_\mu{}^{22}$
satisfy the algebra
\bea
V_{[\mu}{}^{11}V_{\nu]}{}^{22} &=& -\varepsilon_{\mu\nu\rho}\,
f\,V^{\rho\,12} \;,
\nonumber\\
V_{[\mu}{}^{12}V_{\nu]}{}^{11} &=& \ft12\varepsilon_{\mu\nu\rho}\,
f\,V^{\rho\,11} \;,
\nonumber\\
V_{[\mu}{}^{12}V_{\nu]}{}^{22} &=& -\ft12\varepsilon_{\mu\nu\rho}\,
f\,V^{\rho\,22} \;.
\label{alg}
\eea
According to the second equation of~\Ref{quartic1}, they are normalized as
\bea
0&=& V_\mu{}^{11}V^{\mu\,11}  ~=~ V_\mu{}^{22}V^{\mu\,22}
~=~ V_\mu{}^{12}V^{\mu\,11} ~=~ V_\mu{}^{12}V^{\mu\,22}
\;,\nonumber\\
2f^2 &=& V_\mu{}^{11}V^{\mu\,22} ~=~ -2 V_\mu{}^{12}V^{\mu\,12}
\;.
\label{alg2}
\eea
For non-vanishing $f$, these vectors thus form an
orthogonal basis of the three-dimensional spacetime.
In the explicit basis (\ref{basis}), all algebraic relations (\ref{quartic1})
are summarized by (\ref{alg}) and (\ref{alg2}).
For later use, we use the explicit basis (\ref{basis}) to
define (again for non-vanishing $f$) the antisymmetric matrices
\bea
\Omega^{IJ} \equiv \ft12 \epsilon^{ab} \Gamma^I_{a\dot{A}}\Gamma^J_{b\dot{A}}
\;,
\qquad
\Omega_{\dot{A}\dot{B}} \equiv 
\ft12 \epsilon^{ab}
\Gamma^I_{a\dot{A}}\Gamma^I_{b\dot{B}}
\;,
\label{Omegas}
\eea
satisfying
$\Omega_{\dot{A}\dot{B}} \Omega_{\dot{B}\dot{C}} =
-\delta_{\dot{A}\dot{C}}$\, and
$\Omega^{IJ}\Omega^{JK} = -\delta^{IK}$\,.
Group-theoretically, they manifest the fact that under the above branching (\ref{branching})
of the ${\bf 8}_v$ and ${\bf 8}_c$,
there appears another invariant tensor in their respective tensor products.

In contrast, for $f=0$ (i.e.\ $F^{AB}=0$), the first two equations of (\ref{quartic1}) state that
all vectors $V_\mu{}^{AB}$ are null and mutually orthogonal,
i.e.\ they are all proportional according to $V_\mu{}^{AB}=\Lambda^{AB} V_\mu$, with 
a symmetric matrix $\Lambda^{AB}$ of trace 1.
The fourth equation of (\ref{quartic1}) then imposes
\bea
\Lambda^2 &=& \Lambda
\;,
\eea
for the matrix $\Lambda$, i.e.\ this matrix has a single non-vanishing eigenvalue.
We can choose a basis in which 
the only non-vanishing component of $V_\mu{}^{AB}$ is
\bea
V_\mu{}^{11} = V_\mu{}^{AA} =V_\mu 
\;.
\eea
Accordingly we split the index $A=(1,\tilde{a})$,
corresponding to a breaking of $SO(8)$ to $SO(7)$, 
under which the fundamental representations
branch as
\bea
{\bf 8}_s \rightarrow 1+7\;,\quad
{\bf 8}_v \rightarrow  8\;,\quad
{\bf 8}_c \rightarrow 8
\;.
\label{branching_null}
\eea

\subsection{Differential equations}
\label{subsec:diffeq}

In addition to the algebraic relations induced by Fierz identities,
the Killing spinor equations impose a number of 
differential equations on the tensors $F^{AB}$, $V_\mu{}^{AB}$. 
The first of the Killing spinor equations (\ref{KS1}) implies that
\bea
D_{\mu}\, F^{AB} &=&
2g\,A_1^{C[A}\,V_\mu{}_{\vphantom{1}}^{B]C}\;,
\label{diff1}\\
D_\mu\,V_\nu{}^{AB} &=&
2g\,g_{\mu\nu}\,A_1^{C(A}\,F^{B)C}
+2g\,\varepsilon_{\mu\nu\rho}\,
A_1^{C(A}\,V^{\rho\,B)C}
\label{diff2}
\;,
\eea
where the derivative $D_\mu$ now is the full $SO(8)$
covariant derivative. 
From (\ref{diff1}) we thus obtain
in the basis (\ref{basis})
\bea
\partial_{\mu}\, f &=& 
g\, A_1^{ca} \epsilon^{ab}\,V_\mu{}^{bc}
\;,
\label{df}
\eea
or equivalently, using (\ref{alg2}):
$V^{\mu\, ab}\, \partial_{\mu} f =
-2gf^2 \epsilon_{\vphantom{1}}^{c(a}A_1^{b)c} $\,.
In particular, for vanishing $g$, i.e.\ in the ungauged theory, the function $f$ is constant.
From~(\ref{diff2}) we obtain that
\bea
D_{(\mu} V_{\nu)} &=& 0\;,
\eea
i.e.\ the vector $V_{\mu}$ is a Killing vector of the solution.
Equation~(\ref{alg2}) moreover shows that
this vector is either timelike (for $f\not=0$) 
or null (for $f=0$),
in accordance with the expectations.
In the following sections, we will treat the two cases separately.
From (\ref{df}) together with (\ref{alg2}), we find also that
$V^\mu\partial_\mu f=0$, i.e.\ the function $f$ is also constant along the Killing vector field.

Contracting the dilatino equation (\ref{KS2}) with $\Beps^B$ and $\Beps^B\gamma_\nu$,
respectively, gives rise to two equations for the scalar current ${\cal P}_\mu^{Ir}$:
\bea 
0  &=&{} 
  \ft12\Gamma^I_{B\dot{A}}V^{\mu\,AB} \,{\cal P}_\mu^{Ir}
  +g F^{AB}\,A_2^{B\dot{A} r}
\;,
\label{dilatino1}\\
  0  &=&{} 
 - \ft12\Gamma^I_{B\dot{A}} \,F^{AB} \, {\cal P}_\mu^{Ir}
+  \ft12\varepsilon_{\mu\nu\rho} \Gamma^I_{B\dot{A}}
V^{\rho\,AB}\,{\cal P}^{\nu\,Ir}
  + g A_2^{B\dot{A} r}V_\mu{}^{AB}
  \;.
  \label{dilatino2}
\eea
The first equation 
implies in particular (upon contracting with $\Gamma^J_{A\dot{A}}$ and using (\ref{PgP})
and the definition (\ref{A1A2}))
\bea
  V^{\mu} {P}_\mu^{Ir}
  &=&
-g \theta_{\cal KLMN} {\cal V}^{\cal MN}{}_{Ir}\,
\Big(V^{\mu}A_\mu{}^{\cal KL} + 
f\,{\cal V}^{\cal KL}{}_{JK}\,\Omega^{JK}
 \Big)
\;,
\label{VP}
\eea
with $\Omega^{JK}$ from (\ref{Omegas}).
This shows that with the particular (Coulomb type) gauge choice
\bea
V^{\mu}A_{\mu}{}^{\cal KL} &=& 
-f {\cal V}^{\cal KL}{}_{MN}\,\Omega^{MN}
\;,
\nonumber\\
V^{\mu}Q^{IJ}_{\mu} &=& 0 ~=~
V^{\mu}Q^{rs}_{\mu} 
\;,
\label{VQ}
\eea
of vector and $SO(8)$ gauge freedom, all scalar fields are likewise
constant along the Killing vector field
\bea
{\cal L}_{V}\,{\cal S} ~=~ 0\;.
\eea
It requires some more work and the explicit use 
of the duality equation (\ref{duality}) 
to show that as a consequence of equations (\ref{dilatino1}), (\ref{dilatino2})
also the gauge fields are constant along the Killing vector field.
We come back to this in section~\ref{subsec:vectors}.
The gauge fixing (\ref{VQ}) furthermore suggests to split off the $V_\mu$
contribution in the vector fields and introduce new gauge
fields $\hat{A}^{IJ}_\mu$ according to
\bea
A_{\mu}{}^{\cal KL} &=& 
-\frac{1}{4f}\, {\cal V}^{\cal KL}{}_{MN}\,\Omega^{MN}\,V_\mu +  \hat{A} _\mu {}^{\cal KL}
\;,
\label{Ahat}
\eea
for non-vanishing $f$. This split will play an important role in the following.

To summarize, we have shown in this section, that the invariant tensors
$F^{AB}$ and $V_\mu{}^{AB}$ satisfy the algebraic relations (\ref{alg}) and (\ref{alg2})
as a consequence of the Fierz identities of the underlying Killing spinor. 
Moreover, in terms of these tensors, the Killing spinor equations (\ref{KS1}), (\ref{KS2}) 
take the equivalent form (\ref{diff1}), (\ref{diff2}), (\ref{dilatino1}), (\ref{dilatino2}).
In the following we will study these equations in more detail.

\section{Timelike case: general discussion}
\label{sec:general}

In the previous section we have identified the Killing vector field $V_\mu$
among the tensors built from the Killing spinor. Moreover, we have translated all
the constraints imposed by three-dimensional Fierz identities into the algebraic
relations (\ref{alg}), (\ref{alg2}), and the full content of the Killing spinor equations
into the differential equations (\ref{diff1}), (\ref{diff2}) and (\ref{dilatino1}), (\ref{dilatino2}).
In the following, we will study these equations and their consequences in detail.
We first discuss the case $f\not=0$ of a timelike Killing vector $V^\mu V_\mu > 0$\,.
The null case is presented separately in section \ref{sec:null}.

\subsection{Three-dimensional spacetime}
\label{subsec:properties}

Define the complex vector field 
\bea
Z_\mu &\equiv&  X_\mu+\mathrm{i} Y_\mu
~\equiv~
2\,V_\mu^{12}
+\mathrm{i}( V_\mu^{11}-V_\mu^{22})
\;,
\label{VZ}
\eea
with norm given by (\ref{alg2}) as
\bea
\la{norms}
Z_\mu \bar Z^\mu = -2V_\mu V^\mu =  -8 f^2
\;.
\eea
Then equations (\ref{alg}) and (\ref{alg2}) translate into
\bea
V_{[\mu}Z_{\nu]} &=& \mathrm{i} \varepsilon_{\mu\nu\rho}f\,Z^\rho
\;,\qquad
Z_{[\mu} \bar{Z}_{\nu]} = -2\mathrm{i}\varepsilon_{\mu\nu\rho}f\,V^\rho
\;.
\eea
These equations summarize all the algebraic equations
derived in section~\ref{subsec:algeq}.
They encode the fact that the mutually orthogonal vector fields $V_\mu$ and $Z_\mu$ form a 
canonical tetrad for the three-dimensional
spacetime, which we shall employ in the following.
We can choose 
the three-dimensional vielbein as
\bea
E_\mu{}^a &=&
\frac1{2f}\,
(V_\mu,X_\mu,Y_\mu)
\;,
\qquad
E_a{}^\mu ~=~
\frac1{2f}\,
(V^\mu,-X^\mu,-Y^\mu)^T
\;,
\label{EEfull}
\eea
consistently satisfying $E_\mu{}^a E_b{}^\mu=\delta_b^a$ and
\bea
{\rm det}\,E_\mu{}^a
~=~\frac1{8f^3}\,e\varepsilon^{\mu\nu\rho} \, V_\mu X_\nu Y_\rho
~=~ \frac1{4f^2}\,e\, V_\mu   V^\mu 
~=~ e \;.
\eea

Let us now turn to the differential relations of section~\ref{subsec:diffeq}.
We further 
denote by
\bea
A^v\equiv A_1^{11}+A_1^{22}\;,\qquad
A^z\equiv 2A_1^{12}+\mathrm{i} (A_1^{11}-A_1^{22})
\;,
\label{someA}
\eea
certain components of the scalar tensor $A_1^{AB}$
in the basis (\ref{basis}).
Then equation (\ref{df}) takes the form
\bea
\partial_{\mu} f &=& 
\frac{\mathrm{i}g}4\, ( A^{\bar{z}} Z_\mu -A^z \bar{Z}_\mu   )
\;,
\label{dfbasis}
\eea
whereas (\ref{diff2}) induces
\bea
\nabla_\mu\,V_\nu{} &=&
g\,\varepsilon_{\mu\nu\rho}\,
\left(
A^v\,V^{\rho}
+\ft12( A^z \, \bar Z^{\rho}
+A^{\bar z} \, Z^{\rho})
\right)
\;,
\nonumber\\[1ex]
\nabla_\mu Z_\nu &=& 
2\mathrm{i}gf A^z\,g_{\mu\nu}
+g \varepsilon_{\mu\nu\rho}\,
\left(
A^v\,Z^{\rho}+A^z\,V^{\rho}
\right)
-2\mathrm{i} \,{\cal Q}_\mu\,Z_\nu
\;,
\label{dVZ}
\eea
with the $SO(2)$ connection ${\cal Q}_\mu=\frac14 \Omega^{IJ} {\cal Q}^{IJ}_\mu$\,.
Thereby, we have reformulated the first Killing spinor equation (\ref{KS1}).
Let us note that equation (\ref{dfbasis}) follows as a consequence of (\ref{dVZ})
using that the norm of the Killing vector field is given as $V^\mu V_\mu = 4f^2$\,.
Choosing the vielbein as (\ref{EEfull}), equations (\ref{dVZ}) 
precisely encode the spin connection $\omega_\mu{}^{ab}$
\bea
\omega_{\mu}{}^{01}+\mathrm{i}\, \omega_{\mu}{}^{02} &=&
\frac{\mathrm{i}g}{2f}\,\left(
A^z\, V_{\mu}+A^v\, Z_\mu \right) \,,
\nonumber\\
\omega_{\mu}{}^{12}&=&
\frac{g}{4f}\,
\left(2A^v V_\mu
+A^z\,Z_\mu+A^{\bar{z}}\,\bar{Z}_\mu
\right)
+2 {{\cal Q}}_\mu 
\,.
\label{spinconnection}
\eea
Some more computation shows that with this explicit form of the spin connection, the 
Killing spinor equation (\ref{KS1}) can be explicitly integrated to
\bea
\epsilon^A &=& \sqrt{f}\;\{\hat\epsilon,\mathrm{i}\gamma_0\, \hat\epsilon,0,0,0,0,0,0\}
\;,
\eea
where $\hat\epsilon$ is a constant spinor satisfying the projection\footnote{
For the flat gamma matrices (with tangent space indices) we use
the explicit representation
$\gamma_0=\sigma_2, \gamma_1=-\mathrm{i}\sigma_3, 
\gamma_2=-\mathrm{i}\sigma_1$ in terms of Pauli matrices, 
so that $\gamma^{012}=\gamma_{012}=-\mathrm{i}$.}
\bea
(\gamma_0-\gamma_2)\,\hat\epsilon &=& 0\;.
\eea
We have thus reconstructed the Killing spinor from the bilinear tensors.
Note that it also satisfies $V^\mu \partial_\mu \epsilon^A =0$\,.

\subsection{The scalar current}

Equations (\ref{diff1}) and (\ref{diff2}) moreover
determine the mixed components of the $SO(8)$ connection
${\cal Q}^{AB}_\mu$ 
in the basis (\ref{basis}) according to
\bea
{\cal Q}_\mu^{a\tilde{a}} &=& -\frac{g}{f}\,A_1^{\tilde{a}c}\,V_{\mu \, bc}\,\epsilon^{ab}
\;,
\label{Qmixed0}
\eea
leaving unconstrained the others,
which again manifests the breaking of $SO(8)$ down to $SO(2)\times SO(6)$.
Explicitly,
\bea
\hat{\cal Q}_\mu^{1\tilde{a}}+\mathrm{i} \hat{\cal Q}_\mu^{2\tilde{a}}&=& 
-\frac{g}{2f}\,
(A_1^{1\tilde{a}}-\mathrm{i}A_1^{2\tilde{a}})\,\bar Z_\mu 
\;,
\label{Qmixed}
\eea
where according to (\ref{Ahat}) we have defined $\hat{\cal Q}_\mu$
by splitting off its contribution in $V_\mu$\,
(it is a non-trivial consistency check that this contribution 
as induced by (\ref{Qmixed0})
precisely coincides with the assignment of (\ref{VQ})).
Accordingly, with (\ref{VQ}) the remaining components of ${\cal Q}_\mu^{IJ}$
define the $SO(2)\times SO(6)$ connection
\bea
{\cal Q}_{\mu} &=&
- \frac{g}{2f} A^v\,V_\mu + \hat{\cal Q}_{\mu}  \;,
\nonumber\\
{\cal Q}^{\tilde{a}\tilde{b}}_{\mu} &=&
-\frac{g}{128nf} \Gamma^{IJ}_{[\tilde{a}\tilde{b}}\Gamma^{KL}_{12]} 
 \Gamma^{IJKL}_{\dot{A}\dot{B}}\,A_3^{\dot{A}r\dot{B}r}\,V_\mu
 + \hat{\cal Q}^{\tilde{a}\tilde{b}}_{\mu}
\;.
\label{QQ}
\eea

Let us now consider the remaining part ${\cal P}_\mu^{Ir}$ of the 
scalar current. In the basis (\ref{VZ}) of vector fields and using (\ref{VP})
we expand ${\cal P}_\mu^{Ir}$ as
\bea
{\cal P}_\mu^{Ir}&=& 
-\,\frac{g}{2f}\,\epsilon^{ab}\,\Gamma^I_{a\dot{A}}\,A_2^{b\dot{A} r}\,V_\mu
~+~\hat{\cal P}_\mu^{Ir}
\nonumber\\[.5ex]
&=&
-\,\frac{g}{2f}\,\epsilon^{ab}\,\Gamma^I_{a\dot{A}}\,A_2^{b\dot{A} r}\,V_\mu
~+~\frac1{f}\left({\cal P}^{Ir}\,{Z}_\mu
+\overline{{\cal P}^{Ir}}\,\bar{Z}_\mu \right)
\;,
\label{ansatzP}
\eea
with complex components ${\cal P}^{Ir}$, to be determined.
Plugging (\ref{ansatzP}) into (\ref{dilatino1}) after some calculation
(which makes use of the properties (\ref{alg2}) of the vector fields)
leads to the compact eigenvector equation
\bea
\left(\delta^{IJ}-\mathrm{i}\,\Omega^{IJ} \right)\left({\cal P}^{Jr} +\ft{1}{4}\mathrm{i}g\,B_+^{Jr} \right) &=& 
0
\;,
\label{OmegaP}
\eea
where we have defined the following combinations
\bea
B^{Ir}_\pm&=& \mp\frac{\mathrm{i}}2(\Gamma^I_{1\dot{A}}\pm\mathrm{i} \Gamma^I_{2\dot{A}})\,
(A_2^{1\dot{A}r}\pm\mathrm{i}A_2^{2\dot{A}r})
\;,
\eea
of scalar components of the tensor $A_2^{A\dot{A}r}$ in the basis (\ref{basis}).
It is straightforward to verify, that these $B^{Ir}_\pm$ are eigenvectors of $\Omega^{IJ}$ from (\ref{Omegas})
according to $\Omega^{IJ} B^{Jr}_\pm = \pm\mathrm{i} B^{Ir}_\pm$\,.
The general solution to (\ref{OmegaP}) is thus given by setting
\bea
{\cal P}^{Ir} &=& {\cal P}^{Ir}_- - \frac{\mathrm{i}g}{4}\,B_+^{Ir}
\;,
\eea
where ${\cal P}^{Ir}_-$ is an arbitrary eigenvector of $\Omega^{IJ}$
with eigenvalue $-\mathrm{i}$\,. 
The full solution (\ref{ansatzP}) then takes the form
\bea
\hat{{\cal P}}_\mu^{Ir}&=& 
\frac1{f}\,({\cal P}_+^{Ir}\,\bar{Z}_\mu+{\cal P}_-^{Ir}\,{Z}_\mu)
-\frac{\mathrm{i}g}{4f}\,(B_+^{Ir} Z_\mu-B_-^{Ir}\bar{Z}_\mu)
\;,
\label{solutionP}
\eea
where ${\cal P}_+^{Ir}\equiv\overline{{\cal P}_-^{Ir}}$ is an eigenvector of $\Omega^{IJ}$
with eigenvalue $+\mathrm{i}$.
Some further calculation shows that
(\ref{solutionP}) also identically solves equation (\ref{dilatino2}).
The full content of the Killing spinor equations (\ref{KS1}), (\ref{KS2}) is thus contained in 
the form of the spin connection (\ref{spinconnection}) and the solution 
(\ref{ansatzP}), (\ref{solutionP}) for the scalar current.

Of course the solution for the scalar current (\ref{solutionP}) is consistent only 
if in addition this current satisfies the integrability conditions (\ref{intP})--(\ref{intQ2}). 
This severely constrains the choice of the components~${\cal P}^{Ir}_\pm$.
For the gauged theories $g\not=0$ these equations involve the 
non-abelian field strength ${\cal F}_{\mu\nu}{}^{\cal KL}$ which in turn is related
to the scalar current itself by means of the duality equations (\ref{duality}).
The resulting structure thus is rather intricate and will be treated 
in a separate publication~\cite{gauged_progress}. 
In this paper, we will in section~\ref{sec:ungauged} 
explicitly work out the integrability 
conditions in the ungauged case $g=0$.

\subsection{Vector fields and duality}
\label{subsec:vectors}

The vector fields appearing in three-dimensional supergravity are not propagating
but come with a Chern-Simons coupling which relates their field strength to the
scalar fields by means of the duality equation (\ref{duality})
\bea
\theta_{\cal KLMN}\, {\cal F}_{\mu\nu}{}^{\cal MN} &=&
\theta_{\cal KLMN} \,\varepsilon_{\mu\nu\rho}  {\cal V}^{\cal MN}{}_{Ir}\,{\cal P}^{\rho\,Ir}\;.
\label{duality0}
\eea
For the supersymmetric solutions it turned out to be natural
to explicitly split off the vector field components in the direction of the Killing vector field
according to (\ref{Ahat}).
Accordingly, the field strength ${\cal F}_{\mu\nu}{}^{\cal KL}$ is decomposed into
\bea
{\cal F}_{\mu\nu}{}^{\cal KL} &=&
-\hat{\cal D}_{[\mu}\left(\frac1{2f}{\cal V}^{\cal KL}{}_{MN}\Omega^{MN} V_{\nu]}\right)
+ \hat{\cal F}_{\mu\nu}{}^{\cal KL}
\;,
\eea
where $\hat{\cal D}$ refers to the covariant derivative including only the
vector field $\hat{A}_\mu{}^{\cal KL}$ and its 
non-abelian field strength $\hat{\cal F}_{\mu\nu}{}^{\cal KL}$.
The first term on the r.h.s.\ can be evaluated upon using the relations 
(\ref{dVZ}), (\ref{dV}) and the explicit form of (\ref{solutionP}).
Without going into the details of the derivation (which also require the structure of the quadratic
constraints (\ref{quadcon}) on the embedding tensor and will be discussed in a separate publication),
we note that as a final result the duality equation (\ref{duality0}) takes the form
(for non-vanishing $f$)
\bea
\theta_{\cal KLMN}\, \hat{\cal F}_{\mu\nu}{}^{\cal MN} &=&
\frac{g}{2f}\, \theta_{\cal KLMN} \,\varepsilon_{\mu\nu\rho}  
\left(
{\cal V}^{\cal MN}{}_{IJ}\Omega^{IJ}A^v - {\cal V}^{\cal MN}{}_{Ir}\Omega^{IJ}B^v_{Jr}
\right) V^\rho\;,\qquad\quad
\label{dualityhat}
\eea
with $A^v$ from (\ref{someA}) and $B^v_{Ir}\equiv
\Gamma^I_{1\dot{A}}A_2^{1\dot{A}r}+\Gamma^I_{2\dot{A}}A_2^{2\dot{A}r}$.
This equation shows a few remarkable properties. Note first, that all contributions
proportional to ${\cal P}_\pm^{Ir}$ have dropped out from the original equation (\ref{duality0}).
As a result, the r.h.s.\ of (\ref{dualityhat}) has no contributions of order $g^0$
which has important consequences for the ungauged theory, as we will discuss below.
Second, the r.h.s.\ of (\ref{dualityhat}) is entirely proportional to  $\varepsilon_{\mu\nu\rho}V^\rho$.
This implies in particular, that
\bea
V^\mu \hat{\cal F}_{\mu\nu}{}^{\cal KL} &=& 0\;,
\eea
which finally shows that also the vector fields $\hat{A}_\mu{}^{\cal KL}$ are constant 
in the direction of the Killing vector field.

While the full structure of (\ref{dualityhat}) will be analyzed elsewhere, let us 
discuss here its consequences for the ungauged theory.
In the limit $g=0$, all vector fields consistently decouple from the Lagrangian (\ref{L}).
Still the ungauged theory hosts a remnant of the duality equation 
which is given by the unprojected version of equation (\ref{duality0}):
\bea
  {F}_{\mu\nu}{}^{\cal KL} &=&
\varepsilon_{\mu\nu\rho}\,{\cal V}^{\cal KL}{}_{Ir}\, P^{\rho\,Ir}\;,
\label{duality_ab}
\eea
with abelian field strength ${F}_{\mu\nu}{}^{\cal KL}$. 
Even though the vector fields are no longer part
of the action, they can be defined on-shell by means of this equation.
In particular, the Bianchi identities for the field strength ${F}_{\mu\nu}{}^{\cal KL}$ is precisely
equivalent to the scalar field equations of motion (\ref{eqm:scalars}) at $g=0$:
the r.h.s.\ of (\ref{duality_ab}) is the conserved $SO(8,n)$ Noether current of the
ungauged theory.
This represents the standard duality between vectors and scalar fields
in three dimensions.
From equation (\ref{dualityhat}) we see that in this case
\bea
\hat{F}_{\mu\nu}{}^{\cal KL} &=& 0
\;,
\eea
i.e.\ the vector fields $\hat{A}_\mu{}^{\cal KL}$ are locally flat.
In other words, for supersymmetric solutions of the ungauged theory,
using (\ref{Ahat}), 
the duality equation (\ref{duality_ab}) can be explicitly integrated to 
\bea
A_{\mu}{}^{\cal KL} &=& 
-\frac{1}{4f}\, {\cal V}^{\cal KL}{}_{MN}\,\Omega^{MN}\,V_\mu 
\;,
\label{dualA}
\eea
(for non-vanishing $f$) which allows to express the dual vectors directly in terms of the scalar fields.
This is a remarkable property of the supersymmetric solutions; in general, the dual vectors are nonlocal
functions of the scalar fields.
The relation (\ref{dualA}) is of particular importance when discussing a possible higher dimensional origin
of the three-dimensional solutions.
Most compactifications to three dimensions, e.g.\ the heterotic string on a seven-torus
\cite{Sen:1994wr}, lead to a version of the three-dimensional theory which features propagating 
scalar and vector fields. It is only upon dualizing all vectors into scalars, that the $SO(8,n)$
symmetry of the theory becomes manifest and the action takes the compact form (\ref{L}).
Equation (\ref{dualA}) thus gives an explicit formula for the original three-dimensional 
vectors which can then be lifted up to their higher-dimensional ancestors.
We come back to this discussion in the conclusions.

\subsection{Killing Spinor Identities}
\label{sec:KSI}

In this section we will apply the method of Killing spinor identities \cite{Kallosh:1993wx,Bellorin:2005hy} to determine which field equations are satisfied automatically once the Killing spinor equations are solved. As follows immediately from supersymmetry of an action $S$, if the Killing spinor equations are satisfied, the following relations hold
\bea
\sum_{\rm b} \frac{\delta S}{\delta \phi_{\rm b}} \frac{\partial (\delta_\epsilon \phi_{\rm b})}{ \partial \phi_{\rm f}} &=& 0
\;,
\label{KSI}
\eea
where $\phi_{\rm b}$ and $\phi_{\rm f}$ represent the bosonic and fermionic fields, respectively, of the theory. 
We denote the bosonic equations of motion of (\ref{L}) as
\bea
{\cal E}^\mu{}_{\alpha} \equiv \frac{\delta S}{\delta e_\mu{}^\alpha}\;,\qquad 
{\cal E}^\mu{}_{\cal KL} \equiv \frac{\delta S}{\delta A_\mu{}^{\cal KL}} \;,\qquad
{\cal E}_{Ir} \equiv \frac{\delta S}{\delta \Sigma^{Ir}}
\;,
\eea
where the last derivative is taken with respect to a left invariant vector field $\Sigma^{Ir}$ along the coset
manifold $SO(8,n)/(SO(8)\times SO(n))$.
Equations (\ref{KSI}) thus amount to linear relations among these equations.
Specifically, we find with the bosonic supersymmetry transformations given by~\cite{Marcus:1983hb,Nicolai:2001ac}
\be
\begin{array}{rclrcl}
{\cal S}^{-1}\delta_\epsilon {\cal S} &\equ& Y^{Ir}\;\Beps^A\,\GG^I_{A\dA}\chi^{\dA r}\;, &\quad\;
\; \\[1ex]
\delta_\epsilon e_\mu{}^\Ga &\equ& \I \Beps^A\,\Gg^\Ga\psi^A_\mu \;, &
 \; \\[1ex]
\delta_\epsilon A_\mu{}^{\cal KL} &\equ& 
\multicolumn{4}{l}{-\ft12\VV{\cal KL}{IJ}\,\Beps^A\,\GG^{IJ}_{AB}\psi^B_\mu
+ \I\,\VV{\cal KL}{Ir}\,\Beps^A\,\GG^I_{A\dA}\Gg_\mu\chi^{\dA r} \;,}
\end{array}
\ee
that equations (\ref{KSI}) imply the relations
\bea
2 {\cal E}^\mu{}_{\alpha}\, \Beps^A\,\Gg^\Ga  + \I {\cal E}^\mu{}_{\cal KL} \VV{\cal KL}{IJ}\,\Beps^B\,\GG^{IJ}_{BA} &=&0 \;,
\nonumber\\[1ex]
\la{ksi2}
{\cal E}_{Ir}\;\Beps^A\,\GG^I_{A\dA} + \I {\cal E}^\mu{}_{\cal KL} \,\VV{\cal KL}{Ir}\GG^I_{A\dA}\,\Beps^A\,\Gg_\mu &=& 0\;,
\eea
among the bosonic equations of motion.
After contracting these relations with all possible combinations of $\epsilon^B$ and $\gamma_\nu \epsilon^B$
and using the algebraic relations (\ref{alg}), (\ref{alg2}), one finds that for the timelike case (i.e.\ $f\not=0$) 
they imply the following relations
\bea
{\cal E}^{\mu}{}_\alpha V^\alpha &=& f {\cal E}^\mu{}_{\cal KL} \VV{\cal KL}{IJ}\Omega^{IJ}
\;,\qquad
V^{[\mu}{\cal E}^{\nu]}{}_\alpha ~=~ 0\;,
\nonumber\\[.5ex]
2f{\cal E}^\mu{}_{\cal KL} \VV{\cal KL}{Ir} &=& \Omega_{IJ} {\cal E}^{Jr}V^\mu
\;.
\label{KSItime}
\eea
In particular, we see that most of the Einstein equations ${\cal E}^{(\mu}{}_\alpha\, e^{\nu)\alpha}$ 
are indeed satisfied as a consequence of the Killing spinor equations, except for their component in direction $V^\mu V^\nu$
which is proportional to $V_\mu {\cal E}^\mu{}_{\cal KL} \VV{\cal KL}{IJ}\Omega^{IJ}$.
Similarly, the scalar field equation is satisfied only up to a term proportional to $V_\mu {\cal E}^\mu{}_{\cal KL} \VV{\cal KL}{Ir}$.
In order to ensure that a given solution of the Killing spinor equations solves all equations of motion,
we thus have to impose separately the $V_\mu$ component of the duality equation (\ref{duality}).
Specifically, this amounts to imposing equation (\ref{dualityhat}) that we have encountered in the previous section.
For the ungauged theory, this equation is absent, i.e.\ for $f\not=0$ the full set of equations of motion is satisfied as 
a consequence of the Killing spinor equations.
Note however, that the derivation of (\ref{KSI}) has made implicit use of the integrability relations (\ref{intP})--(\ref{intQ2}) of the 
scalar current, i.e.\ in all cases also these integrability relations will have to be imposed on the solution in order to ensure that it satisfies all equations of motion.

\section{Timelike case: the ungauged theory}
\label{sec:ungauged}

From now on, we concentrate on the case of the ungauged theory
(i.e.\ set  $g=0$),
leaving the analysis of the gauged theories for a separate publication~\cite{gauged_progress}.
For $g=0$, the structure of the spin connection (\ref{spinconnection}) 
simplifies drastically and it has only a single non-vanishing
component 
\bea
\omega_\mu{}^{12} &=& 2 Q_\mu\;.
\label{spinc_ungauged}
\eea
The only non-vanishing component of its curvature thus is
$R_{\mu\nu}{}^{12}= 2Q_{\mu\nu}$, such that the  
Ricci tensor and Ricci scalar of the three-dimensional spacetime are given by
\bea
R_{\mu\nu} &=& 
-\frac1{8f^2}\,R\,Z_{(\mu} \bar{Z}_{\nu)}
\;,
\qquad
\quad
R =
\frac1{f}\, \varepsilon_{\mu\nu\rho}\,Q^{\mu\nu}\,V^\rho
\;.
\label{RbyR}
\eea
Let us note that this equation gives rise to
the interesting factorization structure
\bea
R_{\mu\nu}+2\mathrm{i}\,Q_{\mu\nu} &=& 
-\frac1{8f^2}\,R\,Z_{\mu} \bar{Z}_{\nu}
\;.
\eea
The general solution for the scalar current~(\ref{solutionP}) 
in this case takes the short form
\bea
P_\mu^{Ir}&=& 
\frac{1}{f}\,
\left( P_+^{Ir}\,\bar{Z}_\mu+
P_-^{Ir}\,{Z}_\mu
\right)
\;,
\label{PinZ}
\eea
where as before the components $P_+^{Ir}=\overline{P_-^{Ir}}$
are arbitrary eigenvectors of $\Omega^{IJ}$ from (\ref{Omegas})
corresponding to eigenvalues $\pm\mathrm{i}$, respectively.

\subsection{Special coordinates}

Since $V_\mu$ represents a timelike Killing vector field of the solution,
we can choose coordinates such that $V^\mu\partial_\mu = \partial/\partial t$
and no matter field or metric component depends on the time variable~$t$.
For the spatial part of the three-dimensional spacetime, we use coordinates $x^i$, $i=1, 2$\,.
Normalization then implies that $V_\mu=(4f^2,\rho_1,\rho_2)$ with functions $\rho_i$.
According to (\ref{dfbasis}), $f$ is a constant and for simplification
in the following we rescale $t$ such that $f=\frac12$.

From (\ref{dVZ}) we find that $\partial_{[\mu}V_{\nu]}=0$,
hence $\rho_i=\partial_i\rho$ for a function $\rho$ which can be 
absorbed by redefinition (translation) of $t$.
Moreover, the remaining $2\times2$ block in the vielbein (\ref{EEfull}) 
can be brought into conformal gauge.
Thus in these special coordinates the three-dimensional vielbein reduces to
\bea
E_\mu{}^a 
~=~
\left(
\begin{array}{ccc}
1&0&0\\
0&e^\sigma&0\\
0&0&e^\sigma
\end{array}
\right)
\;,
\eea
with the conformal factor $\sigma$\,. Its spin connection is given by
$\omega_\mu{}^{12} = (0,-\partial_2 \sigma, \partial_1\sigma)$\,.
Accordingly, we find for the Ricci tensor
\bea
R_{11} = R_{22}= -\Box\sigma\qquad \Longrightarrow \qquad
R=2e^{-2\sigma} \Box\sigma
\;,
\label{RRR}
\eea
with the two-dimensional flat Laplacian $\Box=\partial^i \partial_i$\,.
From (\ref{spinc_ungauged}) the $SO(2)$ connection is given by $Q_\mu = \frac12 \omega_\mu{}^{12}$\,,
with curvature $Q_{ij}=\frac12 \epsilon_{ij} \Box\sigma$\,.
This shows that if the three-dimensional spacetime is not flat, the $SO(2)$ connection
is necessarily non-vanishing.
In complex coordinates $z=x^1+\mathrm{i}x^2$ and with (\ref{PinZ}),
the full scalar current takes the form
\bea
\la{QPcomplex}
Q_z=-\frac{\mathrm{i}}2\,\partial_z\sigma
\;,\qquad
P_{z}^{Ir} ~=~
 2e^\sigma\,P_-^{Ir}
\;.
\eea
Projecting for this current the integrability relation (\ref{intQ1}) onto its $SO(2)$ part
gives rise to the equation
\bea
\partial_z \partial_{\bar{z}}\,\sigma &=& -2\,e^{2\sigma} P_+^{Ir}P_-^{Ir}
\;.
\label{Boxsigma}
\eea
Putting this together with (\ref{RRR}) shows that $R=P^\mu P_\mu$
in precise agreement with the Einstein equations (\ref{eqm:Einstein}),
as expected from the general analysis of section~\ref{sec:KSI}.

In order to completely determine the solution it remains to solve
the remaining part of the integrability equations (\ref{intP})--(\ref{intQ2}).
Notably, the first one takes the remarkably simple form
\bea
D_{\bar{z}} \left(e^\sigma\,P_-^{Ir}\right) &=& 0\;.
\label{DPhol}
\eea
This is due to the fact that $P_+^{Ir}$ and $P_-^{Ir}$ are eigenvectors 
corresponding to different eigenvalues of $\Omega^{IJ}$ such that
in this coordinate basis the two terms of (\ref{intP}) must vanish separately.
Note however, that the derivative in (\ref{DPhol}) is covariant in that it 
carries the full composite connections $Q^{IJ}_\mu$ and $Q^{rs}_\mu$. 
In general, it is thus not sufficient to choose
$e^\sigma P_-^{Ir}$ to be a holomorphic function of $z$.
The remaining two integrability equations (\ref{intQ1}), (\ref{intQ2})
take the form
\bea
Q^{IJ}_{z\bar{z}} &=& 8 e^{2\sigma} P_{+}^{r[I}P_{-}^{J]r}
\;,
\qquad
Q^{rs}_{z\bar{z}} 
~=~ 8 e^{2\sigma} P_{+}^{I[r}P_{-}^{s]I}
\;.
\label{intQQ0}
\eea
It is straightforward to verify that any solution to (\ref{DPhol}) also
satisfies the scalar equations of motion (\ref{eqm:scalars})
in accordance with the general relations derived in section~\ref{sec:KSI}.

We have thus reduced the construction of supersymmetric solutions to finding 
common solutions to the two-dimensional equations (\ref{Boxsigma})--(\ref{intQQ0}).
In the following, we will exploit the covariantly holomorphic structure 
of (\ref{DPhol}) to study some explicit examples.

\subsection{Explicit solutions}

In this section, we will illustrate the structure of the covariantly holomorphic equation (\ref{DPhol})
by constructing in detail a few explicit solutions and showing that indeed they satisfy the full
set of equations of motion.
 Let us denote by ${\mathbb P}{}_+^{Ii}$ the four 
(normalized and orthogonal) eigenvectors of $\Omega^{IJ}$ from (\ref{Omegas})
with eigenvalues $+\mathrm{i}$, such that
\bea
({\mathbb P}_+^\dagger {\mathbb P}_+)^{ij} &=& \delta^{ij}\;,\nonumber\\
({\mathbb P}_+ {\mathbb P}_+^\dagger)^{IJ} &=& \ft12\, (\delta-\mathrm{i} \Omega)^{IJ}
\;.
\label{PPP}
\eea
Expanding $P^{Ir}_{\pm}$ in terms of these eigenvectors
we write
\bea
P^{Ir}_- &=& \mathbb{P}^{Ii}_- \, \Sigma{}^{ir}
\;,
\label{generalP}
\eea
with coefficients $\Sigma{}^{ir}$. 
We furthermore define the hermitean matrix ${H}^{rs}=(\Sigma^\dagger \Sigma\,)^{rs}$
and the antisymmetric hermitean matrices
\bea
M^{IJ} &=& \ft12( P^{Ir}_{+} P^{Jr}_- - P^{Jr}_{+} P^{Ir}_-) ~=~
\mathrm{i}\, \Im(\mathbb{P}_+ \overline{\Sigma} \Sigma^T\,\mathbb{P}_+^\dagger)^{IJ}
\;, 
\nonumber\\
N^{rs} &=& \ft12( P^{Ir}_{+} P^{Is}_- - P^{Is}_{+} P^{Ir}_-) ~=~
\mathrm{i}\, (\Im {H})^{rs}~=~
{H}^{[rs]}
\;.
\eea
It is easy to check that $[\Omega, M]=0$\,, i.e.\ $M\in \mathfrak{so}(2)\times \mathfrak{so}(6)$\,,
and $\Omega^{IJ}M^{IJ}=-\mathrm{i} {\rm Tr}\,{H}$,
which implies that $\widehat{M}^{IJ}\equiv M^{IJ}+\ft{\mathrm{i}}8\Omega^{IJ} {\rm Tr}\,{H} \in \mathfrak{so}(6)$.
Note that for $n=4$, the matrix $\widehat{M}^{IJ}$ vanishes.
Some further calculation yields
\bea
M^{IJ} P^{J r}_+ &=& 
\ft12 {H}^{rs} P_+^{Is}
\;,
\qquad
\widehat{M}^{IJ} P^{J r}_+ ~=~
\ft12 \left({H}^{rs}-\ft14\delta^{rs}\, {\rm Tr}\,{H}\right) P_+^{Is}
\;.
\eea
Integrability (\ref{intQQ0}) (projected onto its $SO(2)$ part) 
together with (\ref{Boxsigma})
yields the differential equation:
\bea
\mathrm{i}\partial_z \partial_{\bar{z}} \sigma &=& 
\partial_z Q_{\bar{z}}- \partial_{\bar{z}} Q_z  ~=~ 
-2\mathrm{i} e^{2\sigma}\, {\rm Tr}\,{H}
\;,
\label{intSO2}
\eea
which precisely coincides with the Einstein equation~(\ref{eqm:Einstein}).

In the following we will study an explicit ansatz for the scalar current.
We choose the coefficients in (\ref{generalP}) such that
\bea
P^{Ir}_- &=& \mathbb{P}^{Ii}_- \, {{\cal U}}{}^{ir}\, {\zeta(z,\bar{z})}
\;,
\label{Pans}
\eea
with a yet unconstrained complex function $\zeta(z,\bar{z})$ and a constant matrix
 ${\cal U}$, satisfying $({\cal U}^\dagger {\cal U})^{rs}=\delta^{rs}$.
This requires the number of matter multiplets to satisfy
$n\le 4$ and implies $H^{rs}=|\zeta|^2\, \delta^{rs}$\,.
For the $SO(2)\times SO(6)$ connection $Q_\mu^{IJ}$, we make the ansatz
\bea
Q_z^{IJ} &=& 
-\frac{\mathrm{i}}4\,\partial_z\sigma\, \Omega^{IJ}
- \overline{g(z, \bar{z})} \: \widehat{M}^{IJ}
\;,
\label{Qans}
\eea
in accordance with (\ref{QPcomplex}), and set  $Q_\mu^{rs}=0$. 
Then the second equation of (\ref{intQQ0}) is trivially satisfied, whereas the first equation 
in addition to (\ref{intSO2}) gives
the nontrivial differential equation
\bea
\Big\{ \Re[\partial_z ( g |\zeta|^2)]-4e^{2\sigma} |\zeta|^2 \Big\}\; \widehat{M}^{IJ} &=& 0
\;.
\label{int1}
\eea
Finally, it remains to impose the integrability equation (\ref{DPhol}) which yields
\bea
\partial_{\overline{z}} {\rm ln}\, \zeta +\ft54\partial_{\overline{z}} \sigma + \ft18(n-4)\, g \, |\zeta|^2 &=& 0
\;,
\label{int2}
\eea
which (for $n\not=4$) can be solved for $g$. Plugging this solution back into (\ref{int1}) and using (\ref{intSO2}) yields
a differential equation for $|\zeta|^2$ that has the general solution
\bea
|\zeta|^2 &=& e^{-2(n+1)\sigma/n} \,|\chi(z)|^2
\;,
\eea
with an arbitrary holomorphic function $\chi(z)$. Without loss of generality, we can
choose to set
\bea
\zeta &=& 
e^{-(n+1)\sigma/n} \,\chi(z)
\;,
\label{zeta}
\eea
the undetermined phase of $\zeta$ precisely corresponds to an $SO(6)$
gauge transformation. For $n=4$, equation (\ref{int2})
directly induces (\ref{zeta}).
Putting everything together, we find for the scalar current
\bea
Q_z^{IJ} = 
-\Big\{
\frac{\mathrm{i}}4\, \Omega^{IJ}
+\frac{2}{n|\zeta|^2}   \widehat{M}^{IJ} 
\Big\}\;  \partial_z \sigma
\;,
\qquad
P^{Ir}_z = 
2\,\mathbb{P}^{Ii}_- \, {\cal U}^{ir}\, e^{-\sigma/n} \,\chi(z)
\;,
\eea
while $\sigma$ should satisfy the Liouville equation
\bea
\partial_z \partial_{\bar{z}} \sigma &=& 
-2n\, e^{-2\sigma/n} \,|\chi(z)|^2
\;,
\label{boxsigma}
\eea
that descends from (\ref{intSO2}).
Its general solution can be given 
in terms of an arbitrary holomorphic function $\Phi(z)$
\bea
e^\sigma &=& \left(\frac{\sqrt{2} |\chi|\,(1-|\Phi|^2)}{|\Phi'|}\right)^n
\;.
\eea
Defining $\chi(z)=\Psi(z) \Phi'(z)$, the scalar current and the metric 
are thus given by
\bea
P^{Ir}_z &=& 
\frac{\sqrt{2}\, e^{\mathrm{i}\, {\rm arg}\,\Psi(z)}\,\Phi'}{1-|\Phi|^2} \,\,\mathbb{P}^{Ii}_- \, {\cal U}^{ir}
\;,
\nonumber\\
ds^2 &=& dt^2 - 16 \Big( |\Psi|(1-|\Phi|^2)   \Big)^{2n} dz \overline{dz}
\;, 
\label{metric00}
\eea
where $1\le n\le 4$, which summarizes our supersymmetric solution of the $N=8$ theory.
Its curvature is given by
\bea
R &=& -2^{3-n} n\, |\Psi|^{-8}\, |\Phi'|^2\,\left(1-|\Phi|^2 \right)^{-2(n+1)}
\;.
\eea
We can give the solution in more explicit form by explicitly integrating up the scalar current
to the scalar matrix ${\cal S}$.
For simplicity, we restrict to the case $n=4$ while for arbitrary $n$ this can be done 
in precise analogy. For $n=4$ the $SO(6)$ connection vanishes, and the full scalar current is given by
\bea
J_z &\equiv& {\cal S}^{-1}\partial_z{\cal S} ~=~
-\frac{\mathrm{i}}8\, \Omega^{IJ}X^{IJ} \partial_z \sigma
+ \frac{\sqrt{2}\, e^{\mathrm{i}\, {\rm arg}\,\Psi(z)}\,\Phi'}{1-|\Phi|^2} \,\,\mathbb{P}^{Ir}_- \,Y^{Ir} \;,
\label{JJ}
\eea
where we have furthermore chosen ${\cal U}^{ir}=\delta^{ir}$.
The generators appearing in (\ref{JJ}) take the
form of tensor products 
\bea
\ft12\Omega^{IJ}X^{IJ} = 
\left(
\begin{array}{ccc}
0&1&0\\
-1&0&0\\
0&0&0
\end{array}
\right) \otimes I_4
\;,
\quad
\mathbb{P}^{Ir}_-\,Y^{Ir} =
\frac{1}{\sqrt{2}}
\left(
\begin{array}{ccc}
0&0&1\\
0&0& -\mathrm{i}\\
1&-\mathrm{i}&0
\end{array}
\right)  \otimes I_4
\;,
\eea
such that $SO(8,4)$ splits into four copies of $SO(2,1)$.
Some calculation shows that the current $J_z$ takes the matrix form
\bea
J_z &=& {\bf J}_z + H^{-1}\partial_z H + H^{-1} {\bf J}_z H\;,
\label{Jnew}\\[2ex]
&&\qquad\qquad \mbox{with}\quad
H={\rm exp}\Big\{ \frac{\mathrm{i}}{4}\,{\rm log}(\overline{\Psi}/\Psi) \,\Omega^{IJ} X^{IJ} \Big\} \;,\nonumber\\[1ex]
&&\qquad\qquad \mbox{and}\quad
{\bf J}_z =
\frac{\Phi'}{1-|\Phi|^2}\;
\left(
\begin{array}{ccc}
0&\mathrm{i} \overline{\Phi}&-1\\
-\mathrm{i} \overline{\Phi}&0& \mathrm{i}\\
-1&\mathrm{i}&0
\end{array}
\right)
\otimes I_4
\;.
\nonumber
\eea
In particular, this current depends on $\Psi(z)$ only via an $SO(2)$ gauge transformation.
Equation (\ref{Jnew}) can then be explicitly integrated up to yield the matrix form of ${\cal S}$
\bea
{\cal S}\,H^{-1} &=&
\frac1{1-|\Phi|^2}
\left(
\begin{array}{ccc}
1+\ft12(\Phi^2+\overline{\Phi}{}^2) & 2\,\Re\Phi\, \Im\Phi & -2\, \Re\Phi\\
2\,\Re\Phi\, \Im\Phi & 1-\ft12(\Phi^2+\overline{\Phi}{}^2) &-2\, \Im\Phi \\
-2\, \Re\Phi & -2\, \Im\Phi & 1+|\Phi|^2
\end{array}
\right)  \otimes I_4
\;.\qquad\quad
\eea
This form of the scalar fields together with the metric (\ref{metric00}) 
summarizes the supersymmetric solution in terms of two unconstrained
holomorphic functions $\Phi(z)$ and $\Psi(z)$.
One of them could be absorbed by a conformal redefinition of the two-dimensional coordinates.

Finally, for this solution we may evaluate the explicit form of the 
dual vector fields~(\ref{dualA}) which leads to
\bea
A_{t}{}^{IJ} &=& 
- {\cal S}^{I}{}_M {\cal S}^{J}{}_N\,\Omega^{MN}
~=~ -\frac{1+|\Phi|^2}{1-|\Phi|^2} \,\Omega^{IJ}
\;,
\nonumber\\
A_{t}{}^{Ir} &=& 
- {\cal S}^{I}{}_M {\cal S}^{r}{}_N\,\Omega^{MN}
~=~
\frac{\sqrt{2}\,\mathrm{i}}{1-|\Phi|^2}
\left(\,\overline{\Phi}\, \mathbb{P}^{Ir}_+ - \Phi\, \mathbb{P}^{Ir}_-\right)
\;,
\label{dualA0}
\eea
up to a possible flat contribution.
As discussed above, when lifting this solution to a higher-dimensional theory,
e.g.~the heterotic string on a seven-torus, a number of ten-dimensional fields
will be triggered by these three-dimensional vectors.
For example, the Kaluza-Klein vector of the ten-dimensional metric descends to
the three-dimensional vector field components $A_\mu{}^{I1}$, choosing $r=1$
in the second line of (\ref{dualA0}), in accordance with the breaking of $SO(8,n)$
to $SO(7,n-1)$ manifest after reduction.
The components (\ref{dualA0}) thus become part of the ten-dimensional metric.

This concludes our discussion of explicit examples. With a more sophisticated ansatz
for the scalar current
replacing (\ref{Pans}) and (\ref{Qans}), the construction may be generalized to produce
more complicated solutions of the Killing spinor equations.


\section{The null case}
\label{sec:null}

In this last section we consider the case $f=0$, when $F^{AB}=0$ and
the Killing vector $V^\mu$ is a null vector.
From the differential relations (\ref{diff1}), (\ref{diff2}) we obtain in this case
\bea
{\cal Q}_{\mu}^{1\tilde{a}}=0=
A_1^{1\tilde{a}}
\;,
\qquad
D_\mu V_\nu{} = 2g\varepsilon_{\mu\nu\rho} A_1^{11} V^\rho
\;.
\label{DVnull}
\eea
Choosing coordinates such that the Killing vector field
is given by $V^\mu\partial_\mu = \frac{\partial}{\partial v}$,
the metric can then be cast into the general form
\bea
ds^2 &=& F du^2 - H^2 dx^2 + 2 G du\,dv
\;,
\label{nullmetric}
\eea
with functions $F$, $H$, and $G$ which depend on the coordinates $x$ and $u$ only.
Upon computing its Christoffel symbols, the second equation of (\ref{DVnull}) translates into
\bea
\partial_x G &=& 4gA_1^{11} HG
\;,
\eea
which relates the functions $G$ and $H$.
In particular, for $g=0$, we may choose coordinates such that $G=1$\,.
Further fixing of coordinates then allows to also put $H=1$.

In the following, we specify to the ungauged theory.
The dilatino equation (\ref{dilatino2}) shows that in this case
\bea
P^{Ir}_\mu &=& 
P^{Ir}\, V_\mu
\;,
\label{Pnull}
\eea
with yet unspecified components $P^{Ir}$. Thus in particular $P^{Ir}_{[\mu} P^{Js}_{\nu]}=0$
and the integrability equations (\ref{intQ1}), (\ref{intQ2})
show that $Q_{\mu}^{IJ}$ and $Q_\mu^{rs}$ are flat connections.
The remaining integrability condition
$D^{\vphantom{I}}_{[\mu}P^{Ir}_{\nu]}=\partial^{\vphantom{I}}_{[\mu}P^{Ir}_{\nu]}=0$ 
thus states that $P^{Ir}$ in (\ref{Pnull}) is a function of $u$ only, i.e.
\bea
P^{Ir}_\mu &=& 
P^{Ir}(u) \,V_\mu
\;.
\label{Pu}
\eea
With (\ref{Pu}) we have solved all the algebraic and differential
relations that follow from the Killing spinor equations,
as well as the scalar integrability conditions.
In the null case however, this is not sufficient to imply the full set of equations of motion,
as may be confirmed directly from (\ref{ksi2}):
the $V_\mu V_\nu$ component of the Einstein equations remains to be imposed separately.
With (\ref{Pu}), the Einstein equations (\ref{eqm:Einstein}) take the form
\bea
R_{\mu\nu} &=& P^{Ir}_{\mu} P^{Ir}_{\nu}
\;,
\eea
which in particular implies that the Ricci scalar vanishes $R = P^{Ir}_{\mu} P^{\mu\,Ir}= 0$.
Using the explicit metric (\ref{nullmetric}) with $G=H=1$, the Einstein equations yield
\bea
\partial^2_x F &=& 
2 P^{Ir}P^{Ir}
\;,
\eea
thus $F(u,x)=x^2 P^{Ir}(u)P^{Ir}(u) + x R(u) + T(u)$,
with arbitrary functions $R(u)$, $T(u)$.
The latter may be absorbed by
a further redefinition of coordinates.

Summarizing, for $g=0$ we find as the most 
general supersymmetric null solution the pp-wave
\bea
P^{Ir}_\mu &=& 
P^{Ir}(u)\, V_\mu
\;,\nonumber\\[.5ex]
ds^2 &=& 
\left(x^2 P^{Ir}(u)P^{Ir}(u) + x\, R(u)\right)\, du^2 - dx^2 + 2 du\,dv
\;.
\label{Pgnull}
\eea
The scalar current can be integrated up to a scalar matrix
${\cal S}(u)$ depending on $u$ only. In turn, any such matrix gives rise to a current of the form (\ref{Pgnull}).
As in the timelike case,
we may dualize some of the scalar fields back into vectors
by means of (\ref{duality_ab}).
In the null case and with the above metric, this equation takes the explicit form
\bea
  {F}_{xu}{}^{\cal MN} &=&
2{\cal S}^{[{\cal M}}{}_{K}(u){\cal S}^{{\cal N}]}{}_{r}(u)\,{P}^{Kr}(u)\;,
\label{dual_null}
\eea
and can be integrated in $u$ to obtain the gauge field with
a single non-vanishing component $A_x{}^{\cal MN}(u)$\,.

Finally, we can explicitly solve the Killing spinor equations (\ref{KS1}),
to reconstruct the Killing spinor
\bea
\epsilon &=&
\left(
x^2 P^{Ir}(u)P^{Ir}(u) + x\, R(u)\right)^{1/4}\, \hat\epsilon
\;,
\eea
where $\hat\epsilon$ is a constant spinor satisfying 
$\gamma^u\,\hat\epsilon=0$\,.

\section{Conclusions}

In this paper we have taken the first steps to constructing and classifying the supersymmetric
solutions of half-maximal matter-coupled three-dimensional supergravity. We have translated the Killing spinor equations 
into a set of algebraic and differential relations among the bilinear tensors built from
the Killing spinor. This allows to express the spacetime metric and the scalar current in terms of these tensors.
For the ungauged theory, we have reduced the integrability conditions for the current to the covariant
holomorphicity condition (\ref{DPhol}) and constructed particular solutions by choosing appropriate ansaetze.
For the case of a null Killing vector we have given in section~6 the most general solution of the ungauged theory.

An interesting aspect for the solutions found in this model is their possible higher-dimensional origin.
For this it is important to recall that the dimensional reduction to three dimensions generically leads to a theory
with scalar and vector fields of which the latter have to be dualized into scalars in order to bring the action
into the form (\ref{L}). For example, reduction of the heterotic string on a seven-torus leads to a theory
with global symmetry $SO(7,23)$ and vector fields transforming in the vector representation. Only after 
dualizing the 30 vector fields into scalars by means of (\ref{duality_ab})
the full symmetry $SO(8,24)$ is manifest, and the scalars parametrize the coset space
$SO(8,24)/(SO(8)\times SO(24))$. In particular, before dualization 
only an $SO(7)$ subgroup of the full $SO(8)$ $R$-symmetry group is manifest 
under which the fundamental representations branch as
\bea
{\bf 8}_s \rightarrow 8\;,\quad
{\bf 8}_v \rightarrow  7+1\;,\quad
{\bf 8}_c \rightarrow  8
\;.
\label{branchingR}
\eea
In order to lift the above constructed solutions back to higher dimensions, first a number of scalars will have
to be dualized back into the corresponding vector fields. For this, we have derived the explicit formula (\ref{dualA})
for the vector fields (or (\ref{dual_null}) in the null case).
On the other hand, we have seen in the above construction, that the supersymmetric solutions of (\ref{L})  are organized by particular subgroups of the $R$-symmetry group, namely $SO(2)\times SO(6)$ and $SO(7)$ for the solutions with timelike
and null Killing vector, respectively. These will be broken upon singling out particular scalars according to (\ref{branchingR}).
Comparing the branchings (\ref{branching}) and (\ref{branching_null}) to (\ref{branchingR})
allows to identify the common subgroups and shows that the dualization of 30 scalars back into the original vector fields, 
leaves an underlying manifest $U(3)$ and $G_2$ symmetry for the solutions with timelike and null Killing vector, respectively. 
This group structure will have to be studied in more detail in order to systematically address the higher-dimensional origin in this context.

\medskip

What we have presented in this paper is the first systematic approach using the bilinear tensor analysis to the construction of supersymmetric solutions in three-dimensio\-nal supergravities. It naturally suggests a number of further research directions and generalizations. First of all, while in this paper we have restricted the explicit construction of solutions to the ungauged theory, we have shown that large parts of the structure also find their analogue in the full gauged theory. A more detailed analysis of the gauged theory, in which the non-abelian duality between vector and scalar fields plays a key role will be presented elsewhere~\cite{gauged_progress}. 

The three-dimensional case provides an instructive scenario for the interplay of the coset space geometry of the scalar target space with the structure of the Killing spinor equations. These structures appear in a more compact form than in higher-dimensional theories, due to the fact that in three dimensions all dynamical degrees of freedom are accommodated in the scalar sector and higher rank $p$-forms are absent. Yet, a thorough understanding of the three-dimensional case and in particular of the gauged theory will be of importance for the study of the half-maximal supergravity theories in higher dimensions coupled to $n$ vector multiplets whose scalar fields form similar coset spaces. So far, the systematic study of supersymmetric solutions in matter coupled theories has essentially been restricted to the ungauged quarter-maximal theories,
where in four dimensions the scalar target spaces are described by special K\"ahler and quaternionic K\"ahler geometries, see~\cite{Meessen:2006tu,Huebscher:2006mr,Cacciatori:2008ek,Klemm:2009uw}. 
The gaugings of the half-maximal theories~\cite{Schon:2006kz} on the other hand are organized by symmetry groups similar to the ones studied here, such that the solutions of the Killing spinor equations will exhibit similar structures in their scalar sectors, see~$D=4$ \cite{Ellmer:dipl} for some initial discussion. An ultimate goal would be the extension of the present analysis to the maximal (gauged and ungauged) supergravities in the various dimensions, whose scalar target space geometries are given by exceptional coset space sigma models.

Let us finally note that upon taking a proper flat-space limit~\cite{Bergshoeff:2008ix,Bergshoeff:2008bh},
three-dimensional $N=8$ supergravity reduces to the distinct superconformal BLG model of~\cite{Bagger:2007jr,Gustavsson:2007vu}. It would be interesting to study if techniques similar to the ones presented here can be applied to classify the BPS solutions of the BLG theory, in particular, this should relate to the structures found in~\cite{Jeon:2008bx,Jeon:2008zj}.

\bigskip
\bigskip
\bigskip

\subsection*{Acknowledgements}

We would like to thank the ENS de Lyon, Bo\u{g}azi\c{c}i University and
the IMBM of Istanbul for hospitality during the course of this work.
A part of the calculations performed in section~\ref{sec:general}
has been facilitated by use of the computer algebra
system Cadabra~\cite{Peeters:2006kp,Peeters:2007wn}.
The work of N.S.D. and {\"O}.S. is partially supported by the Scientific and
Technological Research Council of Turkey (T{\"U}B\.{I}TAK).
The work of H.S. is supported in part by the Agence
Nationale de la Recherche (ANR).

\bigskip
\bigskip


\section*{Appendix}

\begin{appendix}

\section{$SO(8,n)$ algebra}
\label{app:so}

\subsection{Commutators}

The non-compact algebra $\mathfrak{so}(8,n)={\rm Lie}\,SO(8,n)$ can be described in closed form
in terms of generators $X^{\cal MN}=-X^{\cal NM}$ with commutators
\bea
[X^{\cal MN},X^{\cal KL}] &=& 2\eta^{\cal M[K}X^{\cal L]N}-2\eta^{\cal N[K}X^{L]M}
\;,
\eea
with the $SO(8,n)$ invariant diagonal metric 
$\eta_{\cal MN} = {\rm diag}(\underbrace{1,1,\dots,1}_{8\times},\underbrace{-1,-1,\dots,-1}_{n\times})$.

\noindent
Upon splitting the index ${\cal M}\rightarrow (I,r)$, such that
$\eta^{IJ}=\delta^{IJ}, \eta^{rs}=-\delta^{rs}$, the algebra takes the form
\bea
{}[X^{IJ},X^{KL}] &=& 2\delta^{I[K}X^{L]J}-2\delta^{J[K}X^{L]I}
\;,\qquad
{}[X^{IJ},Y^{Kr}] ~=~ -2\delta^{K[I}Y^{J]r}
\;,\nonumber\\
{}[X^{pq},X^{rs}] &=& 2\delta^{p[r}X^{s]q}-2\delta^{q[r}X^{s]p}
\;,\qquad
{}[X^{rs},Y^{Ip}] ~=~ -2\delta^{p[r}Y^{Is]}
\;,\nonumber\\
{}[Y^{Ir},Y^{Js}] &=& \delta^{IJ}X^{rs}+\delta^{rs}X^{IJ}
\;.
\label{alg8}
\eea
where for comparison to standard conventions we have furthermore redefined 
the generators $X^{rs}\rightarrow - X^{rs}$.
Here, $X^{IJ}$ and $X^{rs}$ denote the compact generators of $SO(8)$ and $SO(n)$ respectively,
the $8n$ noncompact generators are denoted by $Y^{Ir}$.

\subsection{Coset space $SO(8,n)/ \left( SO(8)\times SO(n)\right)$}

The scalar fields describing the $SO(8,n)/ \left( SO(8)\times SO(n)\right)$
coset space sigma model are parametrized by a group element ${\cal S}\in SO(8,n)$ 
evaluated in the fundamental representation,
i.e.\ by an $(8+n)\times(8+n)$ matrix satisfying 
\bea
{\cal S}\,\eta\, {\cal S}^T &=& \eta
\;.
\eea
The coset structure is expressed by the invariance of the theory under local transformations (\ref{coset}).
In the ungauged theory, the scalar current $J_\mu\equiv {\cal S}^{-1}\partial_\mu{\cal S}$ may be decomposed as
\bea
J_\mu &=& \ft12Q_\mu^{IJ} X^{IJ} 
+ \ft12Q_\mu^{rs} X^{rs} + P_\mu^{Ir} Y^{Ir} 
\;,
\label{current0}
\eea
in terms of the generators (\ref{alg8}) and the Lagrangian of the ungauged theory is given by
the $SO(8)\times SO(n)$ invariant combination $\frac14P_\mu^{Ir}P^{\mu\,Ir}$.
The integrability equations
$2\partial_{[\mu}J_{\nu]}+[J_\mu,J_\nu]=0$ induced by the definition (\ref{current0})
translate into
\bea
D_{[\mu} P_{\nu]} &=& 0\;,
\label{integrabilityP}\\
Q^{IJ}_{\mu\nu} &\equiv& 2\partial_{[\mu}Q^{IJ}_{\nu]} + 2 Q^{IK}_{[\mu}Q^{KJ}_{\nu]}
~=~ -2P_{[\mu}^{Ir}P_{\nu]}^{Jr}
\;,
\label{integrabilityPQ}\\
Q^{rs}_{\mu\nu} &\equiv& 2\partial_{[\mu}Q^{rs}_{\nu]} + 2 Q^{rt}_{[\mu}Q^{ts}_{\nu]}
~=~ -2P_{[\mu}^{Ir}P_{\nu]}^{Is}
\;.
\label{integrabilityQrs}
\eea
In the gauged theory, all derivatives are covariant w.r.t.\ a non-abelian gauge group
according to (\ref{current}) and these relations acquire additional contributions proportional to the 
non-abelian field strength, as given in the main text in (\ref{intP})--(\ref{intQ2}).
Let us further note that (\ref{current0}) may be rewritten as
\bea
D_\mu\,{\cal S}^{\cal M}{}_{I}&=&
{\cal S}^{\cal M}{}_{r}\,{ P}_\mu^{Ir}
\;,
\qquad
D_\mu\,{\cal S}^{\cal M}{}_{r}~=~
{\cal S}^{\cal M}{}_{I}\,{ P}_\mu^{Ir}\;,
\label{dV}
\eea
with $D_\mu$ representing the $SO(8)\times SO(n)$ covariant derivative.

\subsection{$SO(8)$ $\Gamma$-matrix identities}

Here, we list a number of identities for the
$SO(8)$ $\Gamma$-matrices, which have proven useful in the calculations of the main text
\bea
\Gamma^{IJ}_{AB} \Gamma^{IJ}_{CD} &=& 16 \delta_{AB}^{CD}
\;,
\nonumber\\[2ex]
\Gamma^{IJ}_{CD} \Gamma^{IJKL}_{AB}
&=&
-8(\Gamma^{KL}_{C(A}\delta_{B)D}-\Gamma^{KL}_{D(A}\delta_{B)C})
+2\delta_{AB}\Gamma^{KL}_{CD}
\;,
\nonumber\\[2ex]
\Gamma^{IJ}_{CD} \Gamma^{KL}_{EF} \Gamma^{IJKL}_{AB}
&=&
-128\,(\delta_{C[E}\delta_{F](A}\delta_{B)D}
-\delta_{D[E}\delta_{F](A}\delta_{B)C})
+32\,\delta_{AB}\delta_{C[E}\delta_{F]D}
\;,
\nonumber\\[2ex]
\Gamma^{IJ}_{AB} \Gamma^{KL}_{CD}
\Gamma^{IJKL}_{\dot{A}\dot{B}} &=&
\Gamma^{IJ}_{[AB} \Gamma^{KL}_{CD]}
\Gamma^{IJKL}_{\dot{A}\dot{B}}
\;,
\nonumber\\[2ex]
\Gamma^{IJ}_{AB}\,\Gamma^{IJKL}_{\dot{A}\dot{B}} &=&
8\,
(\Gamma^{[K}_{A\dot{A}}\Gamma^{L]}_{B\dot{B}}-
\Gamma^{[K}_{B\dot{A}}\Gamma^{L]}_{A\dot{B}})
-2 \delta_{\dot{A}\dot{B}}\Gamma^{KL}_{AB}
\;,
\nonumber\\[2ex]
\Gamma^{IJ}_{AB}\Gamma^{IJK}_{C\dot{A}} &=&
2 \Gamma^{KJ}_{AB}\Gamma^{J}_{C\dot{A}}+
16 \delta_{C[A} \Gamma^K_{B]\dot{A}}
\;.
\eea

\end{appendix}


\providecommand{\href}[2]{#2}\begingroup\raggedright\endgroup

\end{document}